\documentclass[prc,showpacs,preprintnumbers,12pt]{revtex4-2}
\usepackage{graphicx,latexsym,amssymb,amsmath,mathrsfs,color,multirow,mathrsfs,booktabs,ifpdf,epsfig,dcolumn,bm,longtable}
\usepackage{changes}
\usepackage{hyperref}
\hypersetup{colorlinks=true,linkcolor=blue,filecolor=magenta,urlcolor=cyan}
\date{\today}

\begin{document}
\title{Proton \textit{s}-resonance states of $^{12}$C and $^{14,15}$O within the Skyrme Hartree-Fock mean-field framework}
	\author{Nguyen Le Anh}
	\email{anhnl@hcmue.edu.vn}
	\affiliation{Department of Theoretical Physics, Faculty of Physics and Engineering Physics, University of Science, Ho Chi Minh City, Vietnam}
	\affiliation{Vietnam National University, Ho Chi Minh City, Vietnam}
	\affiliation{Department of Physics, Ho Chi Minh City University of Education, 280 An Duong Vuong, District 5, Ho Chi Minh City, Vietnam}
	
	\author{Young-ho Song}
	\email{yhsong@ibs.re.kr}
	\affiliation{Rare Isotope Science Project, Institute for Basic Science, Daejeon 34047, Korea}

	\author{Bui Minh Loc}
	\email{buiminhloc@ibs.re.kr}
	\affiliation{Center for Exotic Nuclear Studies, Institute for Basic Science (IBS), Daejeon 34126, Korea}

\begin{abstract}
The excitation functions of proton elastic scattering on $^{12}$C and $^{14,15}$O nuclei at the energies near the proton-emission threshold are calculated using the Skyrme Hartree-Fock (SHF) in continuum approach. For each excitation function, the first resonance is identified as the $s$-state resonance of the mean-field theory. For $^{15}$O, whose ground-state spin is nonzero, the $s$-state resonance splits into two resonances via the spin-spin component of the optical potential. With a slight adjustment of the strength of central potential, which is obtained from the SHF in continuum approach, the excitation functions of proton elastic scattering for the three nuclei can be explained with high accuracy. The proposed framework can provide a practical method to explain nuclear scattering at the energies near the proton-emission threshold with minimal experimental input.
\end{abstract}

\maketitle

\newpage
\section{INTRODUCTION}
Unbound atomic nuclei, which can be experimentally observed as resonance peaks in the nuclear excitation functions, have been widely studied through resonant elastic scattering at low energy. Experiments of this kind are performed at radioactive beam facilities, such as CERN-ISOLDE (Europe), GANIL (France), GSI (Germany), RIKEN (Japan), and TRIUMF (Canada). The properties of these resonances reveal the coupling of discrete states to a scattering continuum and information about spectroscopic properties of unbound nuclei \cite{johnson20}. Besides, the data collected from these experiments, particularly those of low-lying resonances, are crucial for accurately computing cross sections and nuclear reaction rates in astrophysics, such as radiative-capture and transfer reactions \cite{Descouvemont2020}.

Light nuclei, especially oxygen isotopes, merit further discussion. Both the neutron-rich and proton-rich oxygen isotopes offer an important testing ground for understanding the structure of exotic nuclei. On the proton-rich side, the properties of the ground states and low-lying excited states of unbound nuclei $^{15,16}$F were measured and discussed from proton elastic scattering on proton-rich isotopes $^{14,15}$O at sub-MeV energies \cite{kekelis78, benenson78,peters03,goldberg04,lee07}, and recently \cite{stefan14,degrancey16}. As a light nucleus at the proton drip line, $^{16}$F is an ideal case for studying the structure and reaction of exotic nuclei. In addition, the structure of $^{16}$F has been discussed in the context of isospin-symmetry breaking \cite{fortune06,fortune18,michel22}.

As resonances play an essential role not only in nuclear physics but also in physics in general, theoretical analysis was developed back in the early day of the field with the $R$-matrix theory \cite{Wigner1947, lane58}. Nowadays, the properties of resonances are widely analyzed with the $R$-matrix method, see Refs.~\cite{descouvemont10,azuma10}.
Besides, new theoretical models have been developed. The coupled cluster model \cite{baye05} is specialized for describing the low-lying resonances. The shell model with coupling to the continuum known as the Gamow shell model \cite{hagen06,tsukiyama09,sun17} has been applied with success to reproduce the experimental results including nuclear states with a complicated configuration \cite{michel22}. In the shell-model approach, low-lying resonances near the emission threshold are suggested to be closely related to the configuration of the core and the proton in a single-particle state. In the framework of the mean-field theory for scattering problems (i.e. the optical potential), they can be simply identified as single-particle resonant states.

The optical-potential analysis for nucleon scattering from the target with nonzero spin provides information on spin dependence of the nucleon-nucleus optical potential such as the spin-spin component proposed long ago in Ref.~\cite{Feshbach1958}. It is crucial in the low-energy elastic scattering on light nuclei. As the absorption is absent at the energy of a few MeV, the elastic scattering reflects the nuclear structure as a function of energy, and different spin states are observed in the energy spectrum.
A number of experiments were designed to measure the properties of the spin-spin component \cite{Nagamine1970, Batty1971, Fisher1972, Birchall1974, VonPrzewoski1991, Henneck1994, Hannen2003}. The work in Ref.~\cite{Nagadi2004} discussed that uncertainties still remand in experimental and theoretical analyses because of the experimental precision and complicated nuclear mechanism. This subject was left aside for a while.
Nowadays, as radioactive beam facilities produce more exotic nuclei, a highly accurate optical model potential including spin-dependent term is essential. The microscopic nucleon optical model for target nuclei with nonzero spin within the folding model was intensively re-examined~\cite{Cunningham2013}. Within the folding calculation, the spin dependence of the nucleon-nucleus optical potential is directly linked to that of the effective nucleon-nucleon interaction. The recent study of spin-polarized neutron stars emphasized the necessity of precise measurements for the strength of the spin dependence of effective nucleon-nucleon interactions \cite{Khoa2022}.

In our present work, proton elastic scattering was analyzed using the Skyrme Hartree-Fock (SHF) optical potential, where the SHF calculation is extended for the continuum. The same mean-field potential that originally describes the single-particle bound-state also accounts for the single-particle scattering states \cite{vautherin68, dover71, dover72}. The method of calculation was applied successfully on several low-energy radiative-capture reactions \cite{anh2021PRC103,anh2021PRC104,anh2022PRC}. Recently, the new low-lying narrow resonant state found in the $^{10}$Be($p,p$) scattering \cite{ayyad22, lopez22} was also explained as an $s$-state resonance \cite{anh2022Be10}.

The description of the SHF optical potentials for target nuclei with various ground-state spins-parities is presented in the next section. Our present work is restricted to the $s$-wave scattering. The contribution of the spin-orbit potential vanishes which is convenient for the study of the spin-spin potential. The method was applied on a stable nucleus $^{12}$C and proton-rich oxygen isotopes $^{14,15}$O. The $s$-wave scattering data and resonances are well reproduced with minimal adjustment of the SHF central potential strength. Our result gives information about the properties of the nuclear central potential, especially the spin-spin potential.

\section{Skyrme Hartree-Fock optical potential}
\label{SectionMethod}
As only the $s$-wave is of interest, the scattering equation is written as
\begin{align}\label{Eq:SE1}
 \left(-\dfrac{\hbar^2}{2m}\dfrac{d^2}{dr^2} + \lambda V_{\rm cent} + V_{\rm Coul}\right)\chi_s = E \chi_s.
\end{align}
The nuclear central part $V_{\rm cent}$ and the Coulomb part $V_{\rm Coul}$ of the optical potential are obtained from the SHF calculation \cite{dover71,dover72}:
\begin{align}\label{Eq:Vcent}
 V_{\rm cent} &= \dfrac{m^*}{m} \left[ V^{\rm HF}_{\rm cent} + \dfrac{1}{2}\dfrac{d^2}{dr^2}\dfrac{\hbar^2}{2m^* } -\dfrac{m^*}{2\hbar^2} \left(\dfrac{d}{dr}\dfrac{\hbar^2}{2m^* }\right)^2\right] + \left(1-\dfrac{m^* }{m} \right)E,\\
 V_{\rm Coul} &= \dfrac{m^*}{m} V^{\rm HF}_{\rm Coul}.
\end{align}
The one-body potentials including $V^{\rm HF}_{\rm cent}$ and $V^{\rm HF}_{\rm Coul}$, and the effective mass $m^*$ are given in the SHF formalism as functions of $r$ only. 
One adjustable parameter $\lambda$ for the nuclear central potential in Eq.~\eqref{Eq:SE1} is introduced to reproduce the exact location of the low-energy $s$-state resonance. Note that the optical potential is purely real at this low-energy range. There is a slight energy dependence in $V_{\rm cent}$ caused by the second term in Eq.~\eqref{Eq:Vcent}.

In the case the target has a nonzero spin $I$ as for $^{15}$O($p,p$), the spin-spin interaction plays a role. The optical potential for the target with nonzero spin is
\begin{align} \label{VI=1/2}
V = \lambda V_{\rm cent} + V_{\rm Coul} + V_{sI} ({\bm s} \cdot {\bm I}).
\end{align}
The last term in Eq.~\eqref{VI=1/2} is the spin-spin potential \cite{Feshbach1958,davies64}.
To minimize the number of parameters, the spin-spin potential has the same form factor as the central term \cite{amos2003,amos21}, i.e. $V_{sI} = \alpha V_{\rm cent}$.
The validity of this assumption will be justified by the comparison with the experimental data. Under this simple assumption, the potential is rewritten as \begin{align} \label{Eq:Core-coupling}
	V_{J} = \lambda_J V_{\rm cent} + V_{\rm Coul},
\end{align}
where $\lambda_J\equiv \left[\lambda + \alpha ({\bm s} \cdot {\bm I})\right]$.
In particular, the $^{15}$O($p,p$) reaction has parameters $\lambda_{J=0}=\lambda - \frac{3}{4} \alpha$ for the $0^{-}$ resonance and $\lambda_{J=1}=\lambda + \frac{1}{4} \alpha$ for the $1^-$ resonance in $^{16}$F. The parameters $\lambda_{J=0,1}$ are adjusted according to the experimental energies of the resonances. The final result is obtained by averaging the parameters for the two resonance states with ($2J + 1$) weight. All potentials are deduced from the SHF calculation using the computer program \texttt{skyrme\_rpa} provided in Ref.~\cite{colo13}. The scattering equation is solved by the \texttt{ECIS06} code \cite{ECIS06}. The single-particle width of the $s$-wave resonance is extracted from the inverse of the first derivative of phase shifts $\delta(E)$ at the resonance energy $E_R$ in the center of mass frame
\begin{equation} \label{eq:phase}
\Gamma_{\text{cal}} = 2\left[ \dfrac{d\delta(E)}{dE} \bigg|_{E=E_R} \right]^{-1}.
\end{equation}

\section{Results and discussions}
\subsection{Spinless targets $^{12}$C and $^{14}$O}
\begin{figure}
\centering
\includegraphics[width=\textwidth]{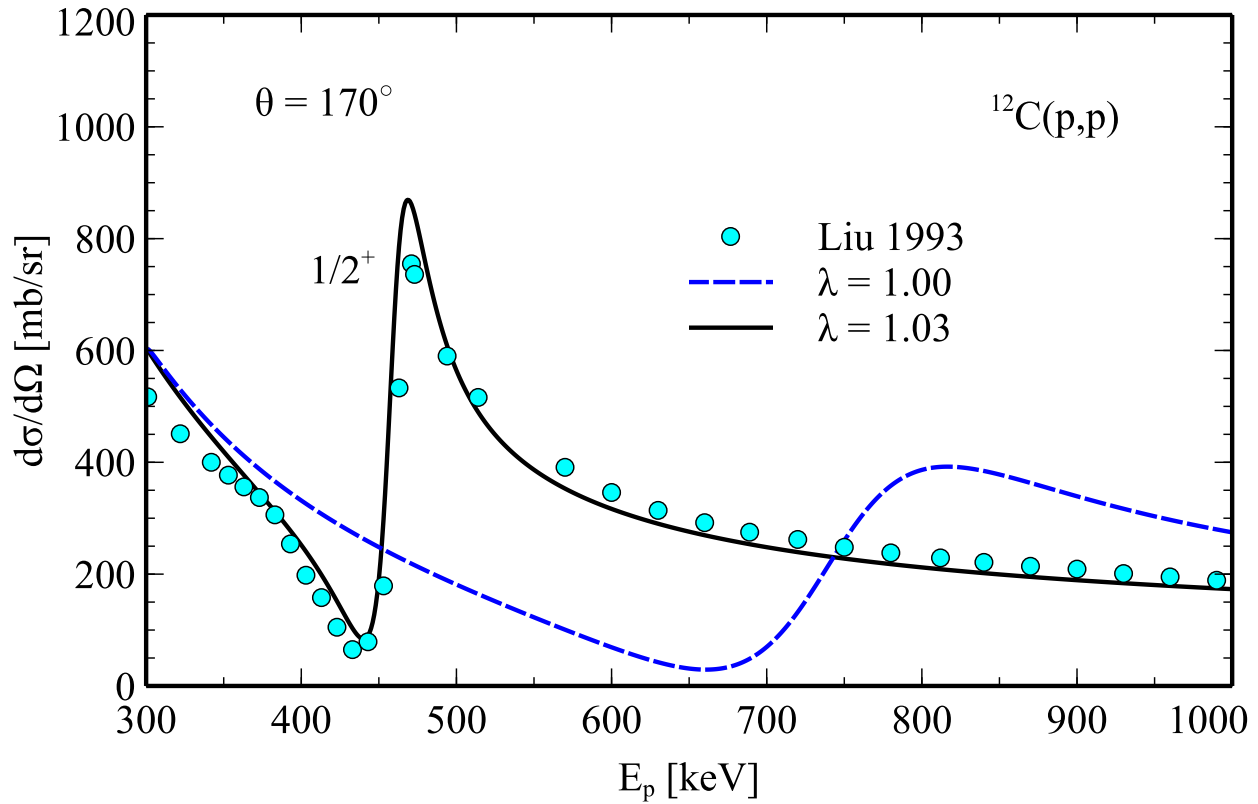}
\caption{The excitation function of the $^{12}$C($p,p$) scattering in the $300$--$1000$ keV range. The resonance at $420$ keV is the $s$-state resonance. The experimental data at the scattering angle of $170^{\circ}$ are taken from Ref.~\cite{liu93}.}
\label{fig:12Cpp}
\end{figure}

\begin{figure}
 \centering
 \includegraphics[width=\textwidth]{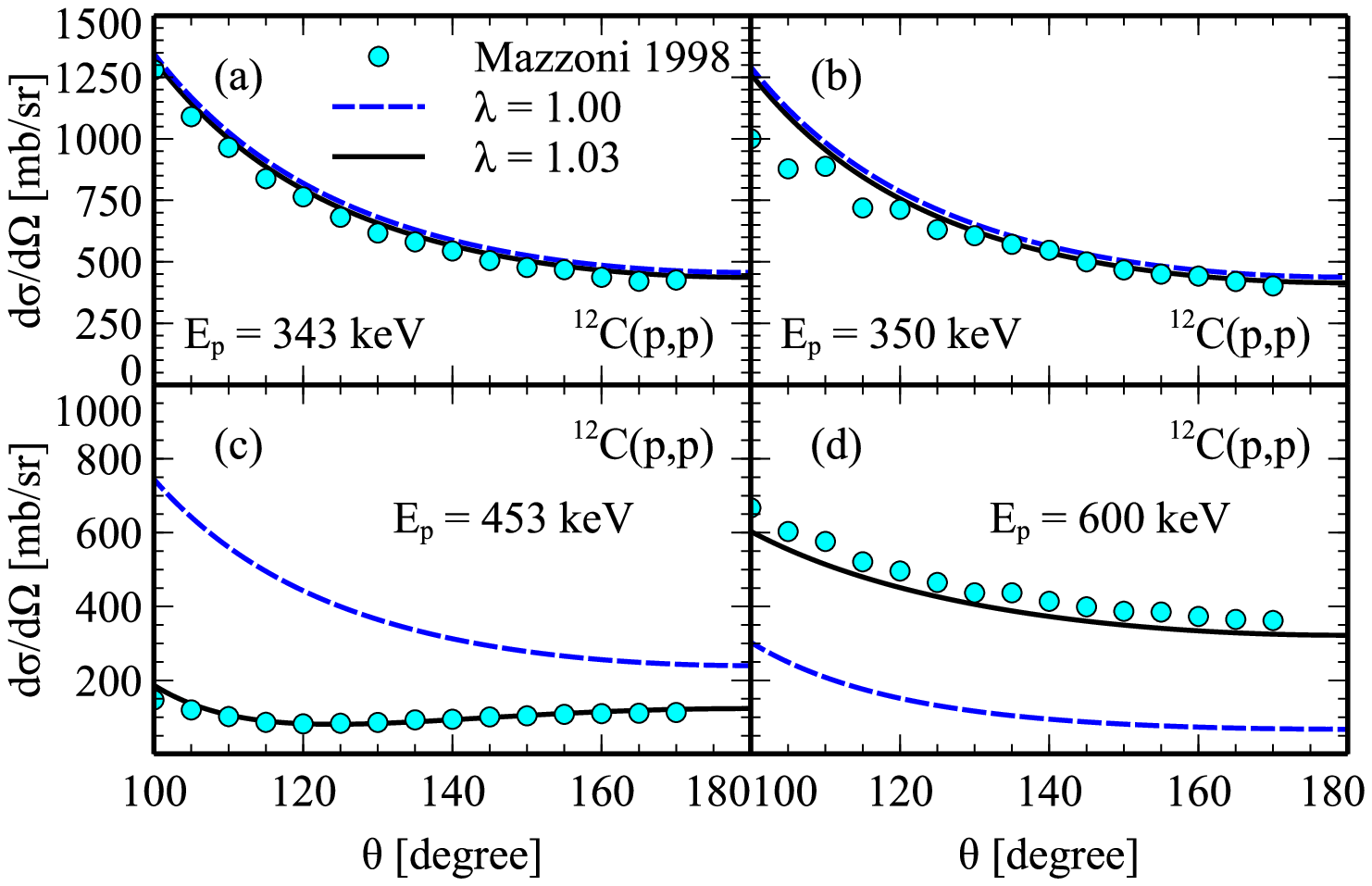}
 \caption{Angular distributions of differential cross section of $^{12}$C($p,p$) around the resonance of $420$ keV. The slight adjustment of the scaling parameter $\lambda$ from $1.00$ to $1.03$ (solid lines) gives the best agreement with literature, especially at $453$ keV. The experimental data are taken from Ref. \cite{mazzoni98}.}
 \label{fig:12Cpp_angle}
\end{figure}

First, the $^{12}$C($p,p$) scattering is considered, which exhibits the first resonance at $420$ keV above the proton-emission threshold of $^{13}$N ($1.94$ MeV). It is in good agreement with the observed resonance with $\Gamma = 31.7$ keV \cite{selove91} at the same energy in the proton radiative-capture reaction.
In the mean-field theory, this resonance is identified as the $s$-state resonance corresponding to the state $1/2^+$ of $^{13}$N at $2.36$ MeV. Fig.~\ref{fig:12Cpp} shows our results with and without the adjustment of the central potential from the SLy4 interaction \cite{chabanat98}. The slight increase of $\lambda$ from $1.00$ to $1.03$ led to a great agreement with the experimental data, especially the shape of the resonance. The SLy4 interaction without the adjustment of $\lambda$ also predicts the existence of the $s$-state resonance; however, the location is not exact as shown in Fig.~\ref{fig:12Cpp}. 
For the purely single-particle state such as this state, the SHF calculation reproduces the energy distribution of the cross section that agrees with the experimental data in Ref.~\cite{liu93} and phenomenological $R$-matrix \cite{Meyer1976}. The angular distribution of the cross section is also obtained with the same potential as shown in Fig.~\ref{fig:12Cpp_angle}.
The necessity of adjustment of the parameter $\lambda$ is significant near the resonance. The calculated angular distribution of the cross section at energies outside the resonance region does not change (Figs.~\ref{fig:12Cpp_angle}(a) and \ref{fig:12Cpp_angle}(b)) while there is a remarkable difference in the vicinity of the resonance (Figs.~\ref{fig:12Cpp_angle}(c) and \ref{fig:12Cpp_angle}(d)). Note that while the reported SHF calculation is based on Ref.~\cite{colo13} which is assumed spherical symmetry and excluded pairing correlations, the calculation without adjustment of the strength of the potential still provides a good description of the nonresonant cross section and hints of possible resonances in the energy range of interest for the investigated nuclei.

\begin{figure}
\centering
\includegraphics[width=\textwidth]{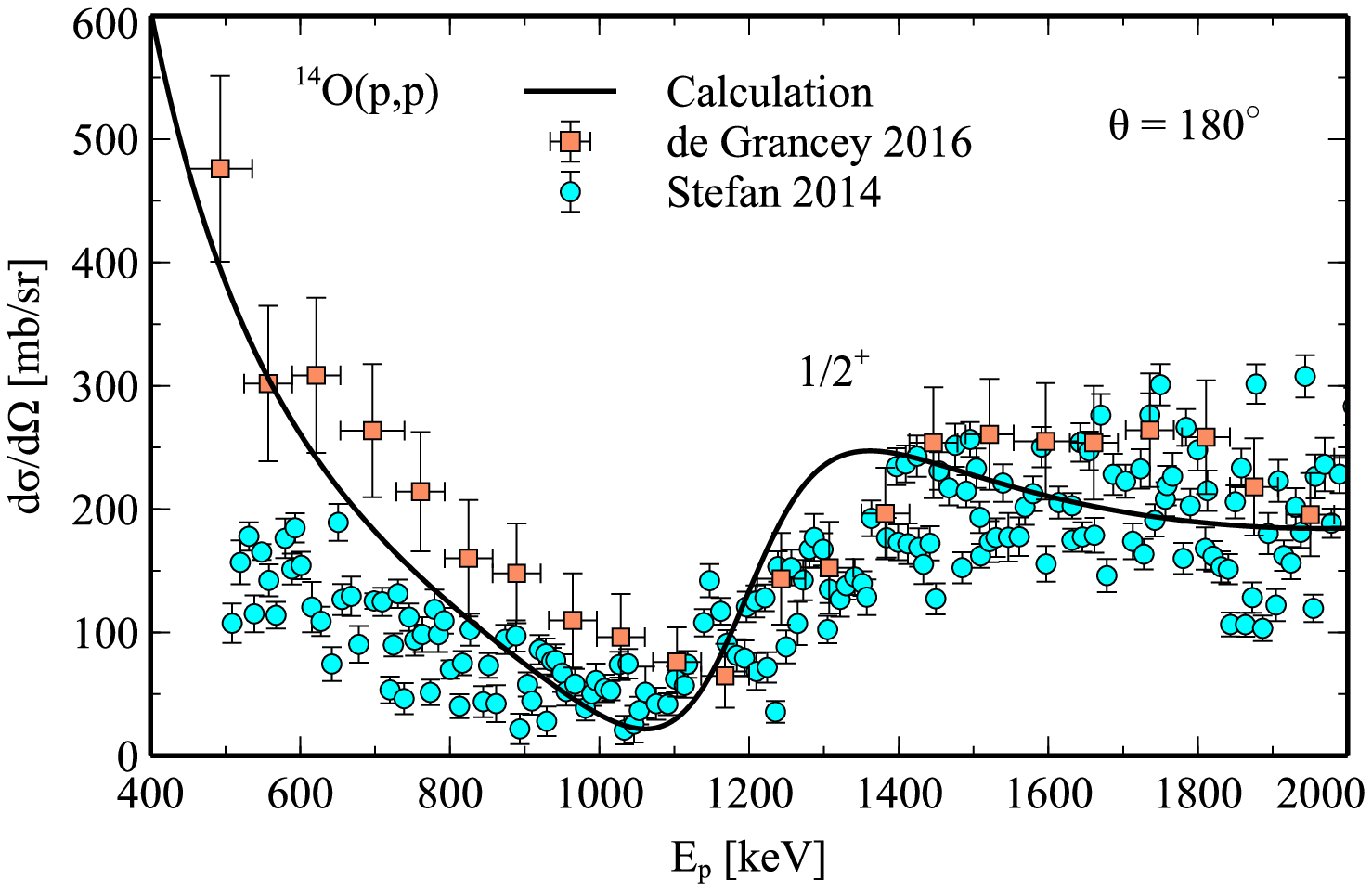}
\caption{Excitation function of $^{14}$O($p,p$) below $2$ MeV. The resonance at $1.27$ MeV is the $s$-state resonance. The calculated different cross section was obtained with $\lambda = 1.02$. The experimental data at $\theta_{\rm lab} = 180^{\circ}$ are from Ref.~\cite{stefan14} (circles) and Ref. \cite{degrancey16} (squares).}
 \label{fig:14Opp}
\end{figure}

The same method can be applied not only for stable nuclei but also for exotic nuclei without additional modifications. The keV-proton elastic scattering from $^{14}$O was first reported in Refs.~\cite{kekelis78,benenson78}. It was then re-measured with higher precision \cite{peters03,stefan14,degrancey16}. The first two resonances in $^{15}$F were described as the single-particle resonances using the Woods-Saxon potential in Ref.~\cite{peters03}. Our calculation confirms that the $1/2^+$ ground state of $^{15}$F is the result of the single-particle resonances of the $s$-scattering wave.
Fig.~\ref{fig:14Opp} presents the calculated differential cross section of the $^{14}$O($p,p$) scattering using the SHF optical potential with $\lambda = 1.02$, which is consistent with the two experimental data sets \cite{stefan14,degrancey16}.

\subsection{Nonzero spin target $^{15}$O and the role of spin-spin potential}
\begin{figure}
 \centering
\includegraphics[width=\textwidth]{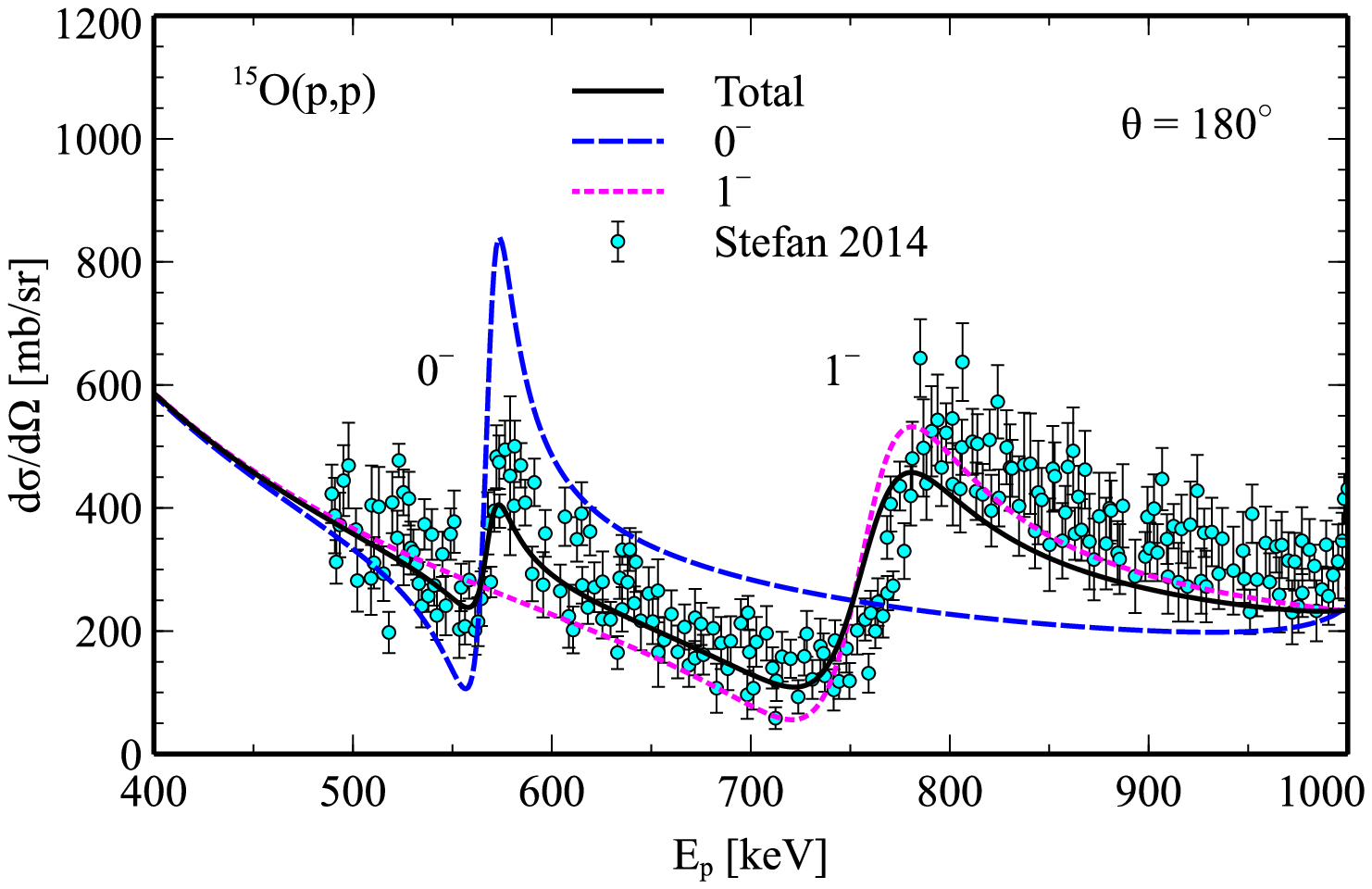}
\caption{Excitation function of $^{15}$O($p,p$) below $1000$ keV with two $s$-state resonances. $\lambda_{J = 0} = 0.99$ and $\lambda_{J = 1} = 0.97$ were adopted for the calculation at the two resonances. The experimental data at $\theta_{\rm lab} = 180^{\circ}$ are from Ref. \cite{stefan14}.}
 \label{fig:15Opp}
\end{figure}

The unbound nucleus $^{16}$F was experimentally examined and discussed in Refs.~\cite{fortune95,lee07,stefan14}. The first two states $0^{-}$ and $1^{-}$ are the $s$-resonance states. To explain this splitting, the spin-spin interaction as described earlier is included.
Indeed, when taking into account the effect of the $^{15}$O target spin of $1/2^{-}$, the $s$-wave scattering generates two resonances with the total spins of $J^\pi = 0^-$ and $J^\pi = 1^-$.
The experimentally observed two resonances are well reproduced by SHF optical potential using $\lambda_{J = 0} = 0.99$ for the first at $0.57$ MeV (dashed line) and $\lambda_{J = 1} = 0.97$ for the second at $0.78$ MeV (dotted line) as shown in Fig.~\ref{fig:15Opp}. 
The values of the parameters show the adjustment of the central potential is 0.98, while the parameter $\alpha$ which describes the relative strength of the spin-spin potential compared to the central potential is $\alpha = 0.02$. Without any adjustment ($\lambda=1,\alpha=0$), the existence of the resonances can still be predicted by SHF optical potential. Although the strength of the spin-spin potential is small, it is responsible for the splitting of the $0^-$ and $1^-$ resonances.

The agreement between the experimental data and SHF calculation points out that the resonant states are in purely single-particle configurations. In many other cases such as $p$-, $d$-states, and even the $s$-resonance state in $^{11}$B in which the $\alpha$-cluster configuration is important \cite{anh2022Be10}, single-particle resonances are mixed with other configurations. Therefore, the $s$-state resonances in the study are the sensitive probes of the single-particle mean-field potential. Furthermore, its splitting is invaluable for studying the spin-spin component of the optical potential.

In the past, the test of the spin-spin potential required sensitive measurements in which the beam was polarized in the depolarized measurements, or the beam and the target were both polarized in the transmission measurements. 
The theoretical formalism for the past experiments was established using the distorted wave Born approximation \cite{davies64}, and the coupled-channel formalism \cite{Tamura1965, stamp67}. Because of the complicated nuclear reaction mechanism, theoretical calculations were significantly improved \cite{Nagamine1970, Hussein1973, MCABEE1990}. The microscopic approach within the folding model using the realistic nucleon-nucleon interaction was applied in Refs.~\cite{Cunningham2011, Cunningham2013}. It has been considered that these studies provided a unique test of the spin dependence in the optical potential.

In our work, the analysis of experimental data measured in Ref.~\cite{stefan14} within the SHF approach provides another reliable method to determine the sign, magnitude, and shape of the spin-spin component of the optical potential. As the elastic scattering was analyzed with the optical potential, the complicated reaction mechanism is completely avoided. The result shows that the strength of the spin-spin potential is 2\% of the central term. It is consistent with the previous conclusions that the strength of spin-spin interaction is of the order of 1 MeV. Although the spin-spin interaction is weak, it is important for understanding its shape. The assumption that the spin-spin potential takes the same form as the central term following Ref.~\cite{amos2003} provides an excellent result in our work.

The calculated widths of resonances are presented in Table \ref{tab:level_parameters} which are not far from the observed width. For the narrow resonance in the stable nucleus $^{12}$C, the location is measured precisely at 420 keV and the width is $31.7 \pm 0.8$ keV \cite{selove91}. The width of 26 keV is computed from Eq.~\eqref{eq:phase} at where the phase shift is $\pi/2$.
The unbound states of $^{16}$F are observed as narrow resonances. The experimental proton-emission widths are given in Ref.~\cite{stefan14} with high accuracy. The calculated widths for $0^-$ and $1^-$ of $^{16}$F are smaller by 10 keV compared to the measurement. The results of narrow widths pointed out that proton-decay channels are mainly dominated in the unbound states of $^{13}$N and $^{15,16}$F.
In the broad resonance of the unbound state of $^{15}$F, the analysis of the data using the potential model gave the width of approximately 700 keV with positions depending on the different definitions for the location of the resonance \cite{goldberg04}. The estimated ground state decay width for $^{15}$F was $800 \pm 300$ keV presented in Refs. \cite{kekelis78,lepine2003}. The relation of the width and the diffuseness parameter within the potential model was discussed carefully in Ref.~\cite{goldberg04}. Our calculation shows the width is 534 or 773 keV corresponding to the definition of resonance position, 1270 or 1400 keV. 

\begin{table}
\setlength{\tabcolsep}{15pt}
    \centering
    \begin{tabular}{ccrrrrr}
        \hline
        Scattering & $J^\pi$ & $E_R$ (c.m.) & $\lambda_J$ & $\Gamma_{\text{cal}}$  & $\Gamma_{\text{exp}}$ \\
                   &         & [keV] &             & [keV] & [keV] \\
        \hline 
        p + $^{12}$C & $1/2^+$ & $420$ & $1.03$ & $26$& $31.7 \pm 0.8$ \cite{selove91} \\
        p + $^{14}$O & $1/2^+$ & $1270$ & $1.02$ & $534$ & $\sim 700$ \cite{goldberg04}\\
         &  & $1400$ & $1.00$ & $773$ & $800 \pm 300$ \cite{kekelis78,lepine2003},\\
         &  & &  &  & $>900$ \cite{benenson78}\\
        p + $^{15}$O & $0^-$ & $536$ & $0.99$ & $15$ & $25.6\pm 4.6$ \cite{stefan14}\\
                     & $1^-$ & $729$ & $0.97$ & $63$& $76 \pm 5$ \cite{stefan14}\\
        \hline
    \end{tabular}
    \caption{Level parameters for the resonances produced in the Skyrme Hartree-Fock approach. The differences between $\Gamma_{\text{cal}}$ and $\Gamma_{\text{exp}}$ are discussed in the text.}
    \label{tab:level_parameters}
\end{table}

Finally, considering the precision of the experimental data used in the analysis, it is worth mentioning the distortion relates to the energy resolution with the radioactive beam. It is about 20 keV in the laboratory frame according to Refs.~\cite{degrancey16,stefan14}. The experimental energy resolution on our results is negligible.

\section{Conclusions and outlooks}
The SHF formalism is a powerful tool for the investigation of not only nuclear structure but also low-energy elastic scattering. With just one parameter, the optical potentials for the elastic scattering of protons on stable and unstable targets $^{12}$C and $^{14,15}$O are obtained from the standard SHF calculation and applied successfully for the resonant elastic scattering in the study of unbound nuclei. The $s$-state resonance and its splitting are sensitive probes of the nuclear central and the spin-spin potential terms, respectively.

In order to improve the method of calculation, one possibility is to use a new Skyrme interaction which is suitable for both bound and unbound nuclei and can be used to study both nuclear structure and scattering. This could be used to further understand how the central potentials of unbound nuclei differ from those of bound nuclei. Additionally, it would be of interest to investigate whether the relatively small magnitude of the spin-spin interaction compared to the central potential, as seen in the $^{15}$O($p,p$) scattering, is a general property for other nuclei. This would provide additional insights into the role of spin-spin interaction in nuclei. In general, future efforts to refine the calculation based on a more microscopic approach are ongoing.

\section*{Acknowledgments}
The authors would like to thank Prof. Naftali Auerbach for the fruitful discussions, Dr. Laszlo Stuhl, and Dr. Jason Park for their discussions and for carefully reading the manuscript. The work of B.M.L. was supported by the Institute for Basic Science (IBS-R031-D1).
The work of Y.H.S. was supported by the Rare Isotope Science Project of Institute for Basic Science, funded by the Ministry of Science and ICT (MSICT) and by the National Research Foundation of Korea (2013M7A1A1075764). 

\bibliography{refs}

\begin{thebibliography}{57}%
\makeatletter
\providecommand \@ifxundefined [1]{%
 \@ifx{#1\undefined}
}%
\providecommand \@ifnum [1]{%
 \ifnum #1\expandafter \@firstoftwo
 \else \expandafter \@secondoftwo
 \fi
}%
\providecommand \@ifx [1]{%
 \ifx #1\expandafter \@firstoftwo
 \else \expandafter \@secondoftwo
 \fi
}%
\providecommand \natexlab [1]{#1}%
\providecommand \enquote  [1]{``#1''}%
\providecommand \bibnamefont  [1]{#1}%
\providecommand \bibfnamefont [1]{#1}%
\providecommand \citenamefont [1]{#1}%
\providecommand \href@noop [0]{\@secondoftwo}%
\providecommand \href [0]{\begingroup \@sanitize@url \@href}%
\providecommand \@href[1]{\@@startlink{#1}\@@href}%
\providecommand \@@href[1]{\endgroup#1\@@endlink}%
\providecommand \@sanitize@url [0]{\catcode `\\12\catcode `\$12\catcode
  `\&12\catcode `\#12\catcode `\^12\catcode `\_12\catcode `\%12\relax}%
\providecommand \@@startlink[1]{}%
\providecommand \@@endlink[0]{}%
\providecommand \url  [0]{\begingroup\@sanitize@url \@url }%
\providecommand \@url [1]{\endgroup\@href {#1}{\urlprefix }}%
\providecommand \urlprefix  [0]{URL }%
\providecommand \Eprint [0]{\href }%
\providecommand \doibase [0]{https://doi.org/}%
\providecommand \selectlanguage [0]{\@gobble}%
\providecommand \bibinfo  [0]{\@secondoftwo}%
\providecommand \bibfield  [0]{\@secondoftwo}%
\providecommand \translation [1]{[#1]}%
\providecommand \BibitemOpen [0]{}%
\providecommand \bibitemStop [0]{}%
\providecommand \bibitemNoStop [0]{.\EOS\space}%
\providecommand \EOS [0]{\spacefactor3000\relax}%
\providecommand \BibitemShut  [1]{\csname bibitem#1\endcsname}%
\let\auto@bib@innerbib\@empty
\bibitem [{\citenamefont {Johnson}\ \emph {et~al.}(2020)\citenamefont
  {Johnson}, \citenamefont {Launey}, \citenamefont {Auerbach}, \citenamefont
  {Bacca}, \citenamefont {Barrett}, \citenamefont {Brune}, \citenamefont
  {Caprio}, \citenamefont {Descouvemont}, \citenamefont {Dickhoff},
  \citenamefont {Elster}, \citenamefont {Fasano}, \citenamefont {Fossez},
  \citenamefont {Hergert}, \citenamefont {Hjorth-Jensen}, \citenamefont
  {Hlophe}, \citenamefont {Hu}, \citenamefont {Betan}, \citenamefont {Idini},
  \citenamefont {König}, \citenamefont {Kravvaris}, \citenamefont {Lee},
  \citenamefont {Lei}, \citenamefont {Mercenne}, \citenamefont {Perez},
  \citenamefont {Nazarewicz}, \citenamefont {Nunes}, \citenamefont
  {P{\l}oszajczak}, \citenamefont {Rotureau}, \citenamefont {Rupak},
  \citenamefont {Shirokov}, \citenamefont {Thompson}, \citenamefont {Vary},
  \citenamefont {Volya}, \citenamefont {Xu}, \citenamefont {Zegers},
  \citenamefont {Zelevinsky},\ and\ \citenamefont {Zhang}}]{johnson20}%
  \BibitemOpen
  \bibfield  {author} {\bibinfo {author} {\bibfnamefont {C.~W.}\ \bibnamefont
  {Johnson}}, \bibinfo {author} {\bibfnamefont {K.~D.}\ \bibnamefont {Launey}},
  \bibinfo {author} {\bibfnamefont {N.}~\bibnamefont {Auerbach}}, \bibinfo
  {author} {\bibfnamefont {S.}~\bibnamefont {Bacca}}, \bibinfo {author}
  {\bibfnamefont {B.~R.}\ \bibnamefont {Barrett}}, \bibinfo {author}
  {\bibfnamefont {C.~R.}\ \bibnamefont {Brune}}, \bibinfo {author}
  {\bibfnamefont {M.~A.}\ \bibnamefont {Caprio}}, \bibinfo {author}
  {\bibfnamefont {P.}~\bibnamefont {Descouvemont}}, \bibinfo {author}
  {\bibfnamefont {W.~H.}\ \bibnamefont {Dickhoff}}, \bibinfo {author}
  {\bibfnamefont {C.}~\bibnamefont {Elster}}, \bibinfo {author} {\bibfnamefont
  {P.~J.}\ \bibnamefont {Fasano}}, \bibinfo {author} {\bibfnamefont
  {K.}~\bibnamefont {Fossez}}, \bibinfo {author} {\bibfnamefont
  {H.}~\bibnamefont {Hergert}}, \bibinfo {author} {\bibfnamefont
  {M.}~\bibnamefont {Hjorth-Jensen}}, \bibinfo {author} {\bibfnamefont
  {L.}~\bibnamefont {Hlophe}}, \bibinfo {author} {\bibfnamefont
  {B.}~\bibnamefont {Hu}}, \bibinfo {author} {\bibfnamefont {R.~M.~I.}\
  \bibnamefont {Betan}}, \bibinfo {author} {\bibfnamefont {A.}~\bibnamefont
  {Idini}}, \bibinfo {author} {\bibfnamefont {S.}~\bibnamefont {König}},
  \bibinfo {author} {\bibfnamefont {K.}~\bibnamefont {Kravvaris}}, \bibinfo
  {author} {\bibfnamefont {D.}~\bibnamefont {Lee}}, \bibinfo {author}
  {\bibfnamefont {J.}~\bibnamefont {Lei}}, \bibinfo {author} {\bibfnamefont
  {A.}~\bibnamefont {Mercenne}}, \bibinfo {author} {\bibfnamefont {R.~N.}\
  \bibnamefont {Perez}}, \bibinfo {author} {\bibfnamefont {W.}~\bibnamefont
  {Nazarewicz}}, \bibinfo {author} {\bibfnamefont {F.~M.}\ \bibnamefont
  {Nunes}}, \bibinfo {author} {\bibfnamefont {M.}~\bibnamefont
  {P{\l}oszajczak}}, \bibinfo {author} {\bibfnamefont {J.}~\bibnamefont
  {Rotureau}}, \bibinfo {author} {\bibfnamefont {G.}~\bibnamefont {Rupak}},
  \bibinfo {author} {\bibfnamefont {A.~M.}\ \bibnamefont {Shirokov}}, \bibinfo
  {author} {\bibfnamefont {I.}~\bibnamefont {Thompson}}, \bibinfo {author}
  {\bibfnamefont {J.~P.}\ \bibnamefont {Vary}}, \bibinfo {author}
  {\bibfnamefont {A.}~\bibnamefont {Volya}}, \bibinfo {author} {\bibfnamefont
  {F.}~\bibnamefont {Xu}}, \bibinfo {author} {\bibfnamefont {R.~G.~T.}\
  \bibnamefont {Zegers}}, \bibinfo {author} {\bibfnamefont {V.}~\bibnamefont
  {Zelevinsky}},\ and\ \bibinfo {author} {\bibfnamefont {X.}~\bibnamefont
  {Zhang}},\ }\href {https://doi.org/10.1088/1361-6471/abb129} {\bibfield
  {journal} {\bibinfo  {journal} {J. Phys. G: Nucl. Part. Phys.}\ }\textbf
  {\bibinfo {volume} {47}},\ \bibinfo {pages} {123001} (\bibinfo {year}
  {2020})}\BibitemShut {NoStop}%
\bibitem [{\citenamefont {Descouvemont}(2020)}]{Descouvemont2020}%
  \BibitemOpen
  \bibfield  {author} {\bibinfo {author} {\bibfnamefont {P.}~\bibnamefont
  {Descouvemont}},\ }\href {https://doi.org/10.3389/fspas.2020.00009}
  {\bibfield  {journal} {\bibinfo  {journal} {Front. Astron. Space Sci.}\
  }\textbf {\bibinfo {volume} {7}},\ \bibinfo {pages} {9} (\bibinfo {year}
  {2020})}\BibitemShut {NoStop}%
\bibitem [{\citenamefont {KeKelis}\ \emph {et~al.}(1978)\citenamefont
  {KeKelis}, \citenamefont {Zisman}, \citenamefont {Scott}, \citenamefont
  {Jahn}, \citenamefont {Vieira}, \citenamefont {Cerny},\ and\ \citenamefont
  {Ajzenberg-Selove}}]{kekelis78}%
  \BibitemOpen
  \bibfield  {author} {\bibinfo {author} {\bibfnamefont {G.~J.}\ \bibnamefont
  {KeKelis}}, \bibinfo {author} {\bibfnamefont {M.~S.}\ \bibnamefont {Zisman}},
  \bibinfo {author} {\bibfnamefont {D.~K.}\ \bibnamefont {Scott}}, \bibinfo
  {author} {\bibfnamefont {R.}~\bibnamefont {Jahn}}, \bibinfo {author}
  {\bibfnamefont {D.~J.}\ \bibnamefont {Vieira}}, \bibinfo {author}
  {\bibfnamefont {J.}~\bibnamefont {Cerny}},\ and\ \bibinfo {author}
  {\bibfnamefont {F.}~\bibnamefont {Ajzenberg-Selove}},\ }\href
  {https://doi.org/10.1103/PhysRevC.17.1929} {\bibfield  {journal} {\bibinfo
  {journal} {Phys. Rev. C}\ }\textbf {\bibinfo {volume} {17}},\ \bibinfo
  {pages} {1929} (\bibinfo {year} {1978})}\BibitemShut {NoStop}%
\bibitem [{\citenamefont {Benenson}\ \emph {et~al.}(1978)\citenamefont
  {Benenson}, \citenamefont {Kashy}, \citenamefont {Ledebuhr}, \citenamefont
  {Pardo}, \citenamefont {Robertson},\ and\ \citenamefont
  {Robinson}}]{benenson78}%
  \BibitemOpen
  \bibfield  {author} {\bibinfo {author} {\bibfnamefont {W.}~\bibnamefont
  {Benenson}}, \bibinfo {author} {\bibfnamefont {E.}~\bibnamefont {Kashy}},
  \bibinfo {author} {\bibfnamefont {A.~G.}\ \bibnamefont {Ledebuhr}}, \bibinfo
  {author} {\bibfnamefont {R.~C.}\ \bibnamefont {Pardo}}, \bibinfo {author}
  {\bibfnamefont {R.~G.~H.}\ \bibnamefont {Robertson}},\ and\ \bibinfo {author}
  {\bibfnamefont {L.~W.}\ \bibnamefont {Robinson}},\ }\href
  {https://doi.org/10.1103/PhysRevC.17.1939} {\bibfield  {journal} {\bibinfo
  {journal} {Phys. Rev. C}\ }\textbf {\bibinfo {volume} {17}},\ \bibinfo
  {pages} {1939} (\bibinfo {year} {1978})}\BibitemShut {NoStop}%
\bibitem [{\citenamefont {Peters}\ \emph {et~al.}(2003)\citenamefont {Peters},
  \citenamefont {Baumann}, \citenamefont {Bazin}, \citenamefont {Brown},
  \citenamefont {Clement}, \citenamefont {Frank}, \citenamefont {Heckman},
  \citenamefont {Luther}, \citenamefont {Nunes}, \citenamefont {Seitz},
  \citenamefont {Stolz}, \citenamefont {Thoennessen},\ and\ \citenamefont
  {Tryggestad}}]{peters03}%
  \BibitemOpen
  \bibfield  {author} {\bibinfo {author} {\bibfnamefont {W.~A.}\ \bibnamefont
  {Peters}}, \bibinfo {author} {\bibfnamefont {T.}~\bibnamefont {Baumann}},
  \bibinfo {author} {\bibfnamefont {D.}~\bibnamefont {Bazin}}, \bibinfo
  {author} {\bibfnamefont {B.~A.}\ \bibnamefont {Brown}}, \bibinfo {author}
  {\bibfnamefont {R.~R.~C.}\ \bibnamefont {Clement}}, \bibinfo {author}
  {\bibfnamefont {N.}~\bibnamefont {Frank}}, \bibinfo {author} {\bibfnamefont
  {P.}~\bibnamefont {Heckman}}, \bibinfo {author} {\bibfnamefont {B.~A.}\
  \bibnamefont {Luther}}, \bibinfo {author} {\bibfnamefont {F.}~\bibnamefont
  {Nunes}}, \bibinfo {author} {\bibfnamefont {J.}~\bibnamefont {Seitz}},
  \bibinfo {author} {\bibfnamefont {A.}~\bibnamefont {Stolz}}, \bibinfo
  {author} {\bibfnamefont {M.}~\bibnamefont {Thoennessen}},\ and\ \bibinfo
  {author} {\bibfnamefont {E.}~\bibnamefont {Tryggestad}},\ }\href
  {https://doi.org/10.1103/PhysRevC.68.034607} {\bibfield  {journal} {\bibinfo
  {journal} {Phys. Rev. C}\ }\textbf {\bibinfo {volume} {68}},\ \bibinfo
  {pages} {034607} (\bibinfo {year} {2003})}\BibitemShut {NoStop}%
\bibitem [{\citenamefont {Goldberg}\ \emph {et~al.}(2004)\citenamefont
  {Goldberg}, \citenamefont {Chubarian}, \citenamefont {Tabacaru},
  \citenamefont {Trache}, \citenamefont {Tribble}, \citenamefont {Aprahamian},
  \citenamefont {Rogachev}, \citenamefont {Skorodumov},\ and\ \citenamefont
  {Tang}}]{goldberg04}%
  \BibitemOpen
  \bibfield  {author} {\bibinfo {author} {\bibfnamefont {V.~Z.}\ \bibnamefont
  {Goldberg}}, \bibinfo {author} {\bibfnamefont {G.~G.}\ \bibnamefont
  {Chubarian}}, \bibinfo {author} {\bibfnamefont {G.}~\bibnamefont {Tabacaru}},
  \bibinfo {author} {\bibfnamefont {L.}~\bibnamefont {Trache}}, \bibinfo
  {author} {\bibfnamefont {R.~E.}\ \bibnamefont {Tribble}}, \bibinfo {author}
  {\bibfnamefont {A.}~\bibnamefont {Aprahamian}}, \bibinfo {author}
  {\bibfnamefont {G.~V.}\ \bibnamefont {Rogachev}}, \bibinfo {author}
  {\bibfnamefont {B.~B.}\ \bibnamefont {Skorodumov}},\ and\ \bibinfo {author}
  {\bibfnamefont {X.~D.}\ \bibnamefont {Tang}},\ }\href
  {https://doi.org/10.1103/PhysRevC.69.031302} {\bibfield  {journal} {\bibinfo
  {journal} {Phys. Rev. C}\ }\textbf {\bibinfo {volume} {69}},\ \bibinfo
  {pages} {031302} (\bibinfo {year} {2004})}\BibitemShut {NoStop}%
\bibitem [{\citenamefont {Lee}\ \emph {et~al.}(2007)\citenamefont {Lee},
  \citenamefont {Per\"aj\"arvi}, \citenamefont {Powell}, \citenamefont
  {O'Neil}, \citenamefont {Moltz}, \citenamefont {Goldberg},\ and\
  \citenamefont {Cerny}}]{lee07}%
  \BibitemOpen
  \bibfield  {author} {\bibinfo {author} {\bibfnamefont {D.~W.}\ \bibnamefont
  {Lee}}, \bibinfo {author} {\bibfnamefont {K.}~\bibnamefont {Per\"aj\"arvi}},
  \bibinfo {author} {\bibfnamefont {J.}~\bibnamefont {Powell}}, \bibinfo
  {author} {\bibfnamefont {J.~P.}\ \bibnamefont {O'Neil}}, \bibinfo {author}
  {\bibfnamefont {D.~M.}\ \bibnamefont {Moltz}}, \bibinfo {author}
  {\bibfnamefont {V.~Z.}\ \bibnamefont {Goldberg}},\ and\ \bibinfo {author}
  {\bibfnamefont {J.}~\bibnamefont {Cerny}},\ }\href
  {https://doi.org/10.1103/PhysRevC.76.024314} {\bibfield  {journal} {\bibinfo
  {journal} {Phys. Rev. C}\ }\textbf {\bibinfo {volume} {76}},\ \bibinfo
  {pages} {024314} (\bibinfo {year} {2007})}\BibitemShut {NoStop}%
\bibitem [{\citenamefont {Stefan}\ \emph {et~al.}(2014)\citenamefont {Stefan},
  \citenamefont {de~Oliveira~Santos}, \citenamefont {Sorlin}, \citenamefont
  {Davinson}, \citenamefont {Lewitowicz}, \citenamefont {Dumitru},
  \citenamefont {Ang\'elique}, \citenamefont {Ang\'elique}, \citenamefont
  {Berthoumieux}, \citenamefont {Borcea}, \citenamefont {Borcea}, \citenamefont
  {Buta}, \citenamefont {Daugas}, \citenamefont {de~Grancey}, \citenamefont
  {Fadil}, \citenamefont {Gr\'evy}, \citenamefont {Kiener}, \citenamefont
  {Lefebvre-Schuhl}, \citenamefont {Lenhardt}, \citenamefont {Mrazek},
  \citenamefont {Negoita}, \citenamefont {Pantelica}, \citenamefont
  {Pellegriti}, \citenamefont {Perrot}, \citenamefont {Ploszajczak},
  \citenamefont {Roig}, \citenamefont {Saint~Laurent}, \citenamefont {Ray},
  \citenamefont {Stanoiu}, \citenamefont {Stodel}, \citenamefont {Tatischeff},\
  and\ \citenamefont {Thomas}}]{stefan14}%
  \BibitemOpen
  \bibfield  {author} {\bibinfo {author} {\bibfnamefont {I.}~\bibnamefont
  {Stefan}}, \bibinfo {author} {\bibfnamefont {F.}~\bibnamefont
  {de~Oliveira~Santos}}, \bibinfo {author} {\bibfnamefont {O.}~\bibnamefont
  {Sorlin}}, \bibinfo {author} {\bibfnamefont {T.}~\bibnamefont {Davinson}},
  \bibinfo {author} {\bibfnamefont {M.}~\bibnamefont {Lewitowicz}}, \bibinfo
  {author} {\bibfnamefont {G.}~\bibnamefont {Dumitru}}, \bibinfo {author}
  {\bibfnamefont {J.~C.}\ \bibnamefont {Ang\'elique}}, \bibinfo {author}
  {\bibfnamefont {M.}~\bibnamefont {Ang\'elique}}, \bibinfo {author}
  {\bibfnamefont {E.}~\bibnamefont {Berthoumieux}}, \bibinfo {author}
  {\bibfnamefont {C.}~\bibnamefont {Borcea}}, \bibinfo {author} {\bibfnamefont
  {R.}~\bibnamefont {Borcea}}, \bibinfo {author} {\bibfnamefont
  {A.}~\bibnamefont {Buta}}, \bibinfo {author} {\bibfnamefont {J.~M.}\
  \bibnamefont {Daugas}}, \bibinfo {author} {\bibfnamefont {F.}~\bibnamefont
  {de~Grancey}}, \bibinfo {author} {\bibfnamefont {M.}~\bibnamefont {Fadil}},
  \bibinfo {author} {\bibfnamefont {S.}~\bibnamefont {Gr\'evy}}, \bibinfo
  {author} {\bibfnamefont {J.}~\bibnamefont {Kiener}}, \bibinfo {author}
  {\bibfnamefont {A.}~\bibnamefont {Lefebvre-Schuhl}}, \bibinfo {author}
  {\bibfnamefont {M.}~\bibnamefont {Lenhardt}}, \bibinfo {author}
  {\bibfnamefont {J.}~\bibnamefont {Mrazek}}, \bibinfo {author} {\bibfnamefont
  {F.}~\bibnamefont {Negoita}}, \bibinfo {author} {\bibfnamefont
  {D.}~\bibnamefont {Pantelica}}, \bibinfo {author} {\bibfnamefont {M.~G.}\
  \bibnamefont {Pellegriti}}, \bibinfo {author} {\bibfnamefont
  {L.}~\bibnamefont {Perrot}}, \bibinfo {author} {\bibfnamefont
  {M.}~\bibnamefont {Ploszajczak}}, \bibinfo {author} {\bibfnamefont
  {O.}~\bibnamefont {Roig}}, \bibinfo {author} {\bibfnamefont {M.~G.}\
  \bibnamefont {Saint~Laurent}}, \bibinfo {author} {\bibfnamefont
  {I.}~\bibnamefont {Ray}}, \bibinfo {author} {\bibfnamefont {M.}~\bibnamefont
  {Stanoiu}}, \bibinfo {author} {\bibfnamefont {C.}~\bibnamefont {Stodel}},
  \bibinfo {author} {\bibfnamefont {V.}~\bibnamefont {Tatischeff}},\ and\
  \bibinfo {author} {\bibfnamefont {J.~C.}\ \bibnamefont {Thomas}},\ }\href
  {https://doi.org/10.1103/PhysRevC.90.014307} {\bibfield  {journal} {\bibinfo
  {journal} {Phys. Rev. C}\ }\textbf {\bibinfo {volume} {90}},\ \bibinfo
  {pages} {014307} (\bibinfo {year} {2014})}\BibitemShut {NoStop}%
\bibitem [{\citenamefont {{de Grancey}}\ \emph {et~al.}(2016)\citenamefont {{de
  Grancey}}, \citenamefont {Mercenne}, \citenamefont {{de Oliveira Santos}},
  \citenamefont {Davinson}, \citenamefont {Sorlin}, \citenamefont
  {Ang\'elique}, \citenamefont {Assi\'e}, \citenamefont {Berthoumieux},
  \citenamefont {Borcea}, \citenamefont {Buta}, \citenamefont {Celikovic},
  \citenamefont {Chudoba}, \citenamefont {Daugas}, \citenamefont {Dumitru},
  \citenamefont {Fadil}, \citenamefont {Gr\'evy}, \citenamefont {Kiener},
  \citenamefont {Lefebvre-Schuhl}, \citenamefont {Michel}, \citenamefont
  {Mrazek}, \citenamefont {Negoita}, \citenamefont {Oko{\l}owicz},
  \citenamefont {Pantelica}, \citenamefont {Pellegriti}, \citenamefont
  {Perrot}, \citenamefont {P{\l}oszajczak}, \citenamefont {Randisi},
  \citenamefont {Ray}, \citenamefont {Roig}, \citenamefont {Rotaru},
  \citenamefont {{Saint Laurent}}, \citenamefont {Smirnova}, \citenamefont
  {Stanoiu}, \citenamefont {Stefan}, \citenamefont {Stodel}, \citenamefont
  {Subotic}, \citenamefont {Tatischeff}, \citenamefont {Thomas}, \citenamefont
  {Uji\'c},\ and\ \citenamefont {Wolski}}]{degrancey16}%
  \BibitemOpen
  \bibfield  {author} {\bibinfo {author} {\bibfnamefont {F.}~\bibnamefont {{de
  Grancey}}}, \bibinfo {author} {\bibfnamefont {A.}~\bibnamefont {Mercenne}},
  \bibinfo {author} {\bibfnamefont {F.}~\bibnamefont {{de Oliveira Santos}}},
  \bibinfo {author} {\bibfnamefont {T.}~\bibnamefont {Davinson}}, \bibinfo
  {author} {\bibfnamefont {O.}~\bibnamefont {Sorlin}}, \bibinfo {author}
  {\bibfnamefont {J.~C.}\ \bibnamefont {Ang\'elique}}, \bibinfo {author}
  {\bibfnamefont {M.}~\bibnamefont {Assi\'e}}, \bibinfo {author} {\bibfnamefont
  {E.}~\bibnamefont {Berthoumieux}}, \bibinfo {author} {\bibfnamefont
  {R.}~\bibnamefont {Borcea}}, \bibinfo {author} {\bibfnamefont
  {A.}~\bibnamefont {Buta}}, \bibinfo {author} {\bibfnamefont {I.}~\bibnamefont
  {Celikovic}}, \bibinfo {author} {\bibfnamefont {V.}~\bibnamefont {Chudoba}},
  \bibinfo {author} {\bibfnamefont {J.~M.}\ \bibnamefont {Daugas}}, \bibinfo
  {author} {\bibfnamefont {G.}~\bibnamefont {Dumitru}}, \bibinfo {author}
  {\bibfnamefont {M.}~\bibnamefont {Fadil}}, \bibinfo {author} {\bibfnamefont
  {S.}~\bibnamefont {Gr\'evy}}, \bibinfo {author} {\bibfnamefont
  {J.}~\bibnamefont {Kiener}}, \bibinfo {author} {\bibfnamefont
  {A.}~\bibnamefont {Lefebvre-Schuhl}}, \bibinfo {author} {\bibfnamefont
  {N.}~\bibnamefont {Michel}}, \bibinfo {author} {\bibfnamefont
  {J.}~\bibnamefont {Mrazek}}, \bibinfo {author} {\bibfnamefont
  {F.}~\bibnamefont {Negoita}}, \bibinfo {author} {\bibfnamefont
  {J.}~\bibnamefont {Oko{\l}owicz}}, \bibinfo {author} {\bibfnamefont
  {D.}~\bibnamefont {Pantelica}}, \bibinfo {author} {\bibfnamefont
  {M.}~\bibnamefont {Pellegriti}}, \bibinfo {author} {\bibfnamefont
  {L.}~\bibnamefont {Perrot}}, \bibinfo {author} {\bibfnamefont
  {M.}~\bibnamefont {P{\l}oszajczak}}, \bibinfo {author} {\bibfnamefont
  {G.}~\bibnamefont {Randisi}}, \bibinfo {author} {\bibfnamefont
  {I.}~\bibnamefont {Ray}}, \bibinfo {author} {\bibfnamefont {O.}~\bibnamefont
  {Roig}}, \bibinfo {author} {\bibfnamefont {F.}~\bibnamefont {Rotaru}},
  \bibinfo {author} {\bibfnamefont {M.~G.}\ \bibnamefont {{Saint Laurent}}},
  \bibinfo {author} {\bibfnamefont {N.}~\bibnamefont {Smirnova}}, \bibinfo
  {author} {\bibfnamefont {M.}~\bibnamefont {Stanoiu}}, \bibinfo {author}
  {\bibfnamefont {I.}~\bibnamefont {Stefan}}, \bibinfo {author} {\bibfnamefont
  {C.}~\bibnamefont {Stodel}}, \bibinfo {author} {\bibfnamefont
  {K.}~\bibnamefont {Subotic}}, \bibinfo {author} {\bibfnamefont
  {V.}~\bibnamefont {Tatischeff}}, \bibinfo {author} {\bibfnamefont {J.~C.}\
  \bibnamefont {Thomas}}, \bibinfo {author} {\bibfnamefont {P.}~\bibnamefont
  {Uji\'c}},\ and\ \bibinfo {author} {\bibfnamefont {R.}~\bibnamefont
  {Wolski}},\ }\href {https://doi.org/10.1016/j.physletb.2016.04.051}
  {\bibfield  {journal} {\bibinfo  {journal} {Phys. Lett. B}\ }\textbf
  {\bibinfo {volume} {758}},\ \bibinfo {pages} {26} (\bibinfo {year}
  {2016})}\BibitemShut {NoStop}%
\bibitem [{\citenamefont {Fortune}(2006)}]{fortune06}%
  \BibitemOpen
  \bibfield  {author} {\bibinfo {author} {\bibfnamefont {H.~T.}\ \bibnamefont
  {Fortune}},\ }\href {https://doi.org/10.1103/PhysRevC.74.054310} {\bibfield
  {journal} {\bibinfo  {journal} {Phys. Rev. C}\ }\textbf {\bibinfo {volume}
  {74}},\ \bibinfo {pages} {054310} (\bibinfo {year} {2006})}\BibitemShut
  {NoStop}%
\bibitem [{\citenamefont {Fortune}(2018)}]{fortune18}%
  \BibitemOpen
  \bibfield  {author} {\bibinfo {author} {\bibfnamefont {H.~T.}\ \bibnamefont
  {Fortune}},\ }\href {https://doi.org/10.1103/PhysRevC.97.044314} {\bibfield
  {journal} {\bibinfo  {journal} {Phys. Rev. C}\ }\textbf {\bibinfo {volume}
  {97}},\ \bibinfo {pages} {044314} (\bibinfo {year} {2018})}\BibitemShut
  {NoStop}%
\bibitem [{\citenamefont {Michel}\ \emph {et~al.}(2022)\citenamefont {Michel},
  \citenamefont {Li}, \citenamefont {Ru},\ and\ \citenamefont
  {Zuo}}]{michel22}%
  \BibitemOpen
  \bibfield  {author} {\bibinfo {author} {\bibfnamefont {N.}~\bibnamefont
  {Michel}}, \bibinfo {author} {\bibfnamefont {J.~G.}\ \bibnamefont {Li}},
  \bibinfo {author} {\bibfnamefont {L.~H.}\ \bibnamefont {Ru}},\ and\ \bibinfo
  {author} {\bibfnamefont {W.}~\bibnamefont {Zuo}},\ }\href
  {https://doi.org/10.1103/PhysRevC.106.L011301} {\bibfield  {journal}
  {\bibinfo  {journal} {Phys. Rev. C}\ }\textbf {\bibinfo {volume} {106}},\
  \bibinfo {pages} {L011301} (\bibinfo {year} {2022})}\BibitemShut {NoStop}%
\bibitem [{\citenamefont {Wigner}\ and\ \citenamefont
  {Eisenbud}(1947)}]{Wigner1947}%
  \BibitemOpen
  \bibfield  {author} {\bibinfo {author} {\bibfnamefont {E.~P.}\ \bibnamefont
  {Wigner}}\ and\ \bibinfo {author} {\bibfnamefont {L.}~\bibnamefont
  {Eisenbud}},\ }\href {https://doi.org/10.1103/PhysRev.72.29} {\bibfield
  {journal} {\bibinfo  {journal} {Phys. Rev.}\ }\textbf {\bibinfo {volume}
  {72}},\ \bibinfo {pages} {29} (\bibinfo {year} {1947})}\BibitemShut {NoStop}%
\bibitem [{\citenamefont {Lane}\ and\ \citenamefont {Thomas}(1958)}]{lane58}%
  \BibitemOpen
  \bibfield  {author} {\bibinfo {author} {\bibfnamefont {A.~M.}\ \bibnamefont
  {Lane}}\ and\ \bibinfo {author} {\bibfnamefont {R.~G.}\ \bibnamefont
  {Thomas}},\ }\href {https://doi.org/10.1103/RevModPhys.30.257} {\bibfield
  {journal} {\bibinfo  {journal} {Rev. Mod. Phys.}\ }\textbf {\bibinfo {volume}
  {30}},\ \bibinfo {pages} {257} (\bibinfo {year} {1958})}\BibitemShut
  {NoStop}%
\bibitem [{\citenamefont {Descouvemont}\ and\ \citenamefont
  {Baye}(2010)}]{descouvemont10}%
  \BibitemOpen
  \bibfield  {author} {\bibinfo {author} {\bibfnamefont {P.}~\bibnamefont
  {Descouvemont}}\ and\ \bibinfo {author} {\bibfnamefont {D.}~\bibnamefont
  {Baye}},\ }\href {https://doi.org/10.1088/0034-4885/73/3/036301} {\bibfield
  {journal} {\bibinfo  {journal} {Rep. Prog. Phys.}\ }\textbf {\bibinfo
  {volume} {73}},\ \bibinfo {pages} {036301} (\bibinfo {year}
  {2010})}\BibitemShut {NoStop}%
\bibitem [{\citenamefont {Azuma}\ \emph {et~al.}(2010)\citenamefont {Azuma},
  \citenamefont {Uberseder}, \citenamefont {Simpson}, \citenamefont {Brune},
  \citenamefont {Costantini}, \citenamefont {de~Boer}, \citenamefont
  {G\"orres}, \citenamefont {Heil}, \citenamefont {LeBlanc}, \citenamefont
  {Ugalde},\ and\ \citenamefont {Wiescher}}]{azuma10}%
  \BibitemOpen
  \bibfield  {author} {\bibinfo {author} {\bibfnamefont {R.~E.}\ \bibnamefont
  {Azuma}}, \bibinfo {author} {\bibfnamefont {E.}~\bibnamefont {Uberseder}},
  \bibinfo {author} {\bibfnamefont {E.~C.}\ \bibnamefont {Simpson}}, \bibinfo
  {author} {\bibfnamefont {C.~R.}\ \bibnamefont {Brune}}, \bibinfo {author}
  {\bibfnamefont {H.}~\bibnamefont {Costantini}}, \bibinfo {author}
  {\bibfnamefont {R.~J.}\ \bibnamefont {de~Boer}}, \bibinfo {author}
  {\bibfnamefont {J.}~\bibnamefont {G\"orres}}, \bibinfo {author}
  {\bibfnamefont {M.}~\bibnamefont {Heil}}, \bibinfo {author} {\bibfnamefont
  {P.~J.}\ \bibnamefont {LeBlanc}}, \bibinfo {author} {\bibfnamefont
  {C.}~\bibnamefont {Ugalde}},\ and\ \bibinfo {author} {\bibfnamefont
  {M.}~\bibnamefont {Wiescher}},\ }\href
  {https://doi.org/10.1103/PhysRevC.81.045805} {\bibfield  {journal} {\bibinfo
  {journal} {Phys. Rev. C}\ }\textbf {\bibinfo {volume} {81}},\ \bibinfo
  {pages} {045805} (\bibinfo {year} {2010})}\BibitemShut {NoStop}%
\bibitem [{\citenamefont {Baye}\ \emph {et~al.}(2005)\citenamefont {Baye},
  \citenamefont {Descouvemont},\ and\ \citenamefont {Leo}}]{baye05}%
  \BibitemOpen
  \bibfield  {author} {\bibinfo {author} {\bibfnamefont {D.}~\bibnamefont
  {Baye}}, \bibinfo {author} {\bibfnamefont {P.}~\bibnamefont {Descouvemont}},\
  and\ \bibinfo {author} {\bibfnamefont {F.}~\bibnamefont {Leo}},\ }\href
  {https://doi.org/10.1103/PhysRevC.72.024309} {\bibfield  {journal} {\bibinfo
  {journal} {Phys. Rev. C}\ }\textbf {\bibinfo {volume} {72}},\ \bibinfo
  {pages} {024309} (\bibinfo {year} {2005})}\BibitemShut {NoStop}%
\bibitem [{\citenamefont {Hagen}\ \emph {et~al.}(2006)\citenamefont {Hagen},
  \citenamefont {Hjorth-Jensen},\ and\ \citenamefont {Michel}}]{hagen06}%
  \BibitemOpen
  \bibfield  {author} {\bibinfo {author} {\bibfnamefont {G.}~\bibnamefont
  {Hagen}}, \bibinfo {author} {\bibfnamefont {M.}~\bibnamefont
  {Hjorth-Jensen}},\ and\ \bibinfo {author} {\bibfnamefont {N.}~\bibnamefont
  {Michel}},\ }\href {https://doi.org/10.1103/PhysRevC.73.064307} {\bibfield
  {journal} {\bibinfo  {journal} {Phys. Rev. C}\ }\textbf {\bibinfo {volume}
  {73}},\ \bibinfo {pages} {064307} (\bibinfo {year} {2006})}\BibitemShut
  {NoStop}%
\bibitem [{\citenamefont {Tsukiyama}\ \emph {et~al.}(2009)\citenamefont
  {Tsukiyama}, \citenamefont {Hjorth-Jensen},\ and\ \citenamefont
  {Hagen}}]{tsukiyama09}%
  \BibitemOpen
  \bibfield  {author} {\bibinfo {author} {\bibfnamefont {K.}~\bibnamefont
  {Tsukiyama}}, \bibinfo {author} {\bibfnamefont {M.}~\bibnamefont
  {Hjorth-Jensen}},\ and\ \bibinfo {author} {\bibfnamefont {G.}~\bibnamefont
  {Hagen}},\ }\href {https://doi.org/10.1103/PhysRevC.80.051301} {\bibfield
  {journal} {\bibinfo  {journal} {Phys. Rev. C}\ }\textbf {\bibinfo {volume}
  {80}},\ \bibinfo {pages} {051301} (\bibinfo {year} {2009})}\BibitemShut
  {NoStop}%
\bibitem [{\citenamefont {Sun}\ \emph {et~al.}(2017)\citenamefont {Sun},
  \citenamefont {Wu}, \citenamefont {Zhao}, \citenamefont {Hu}, \citenamefont
  {Dai},\ and\ \citenamefont {Xu}}]{sun17}%
  \BibitemOpen
  \bibfield  {author} {\bibinfo {author} {\bibfnamefont {Z.~H.}\ \bibnamefont
  {Sun}}, \bibinfo {author} {\bibfnamefont {Q.}~\bibnamefont {Wu}}, \bibinfo
  {author} {\bibfnamefont {Z.~H.}\ \bibnamefont {Zhao}}, \bibinfo {author}
  {\bibfnamefont {B.~S.}\ \bibnamefont {Hu}}, \bibinfo {author} {\bibfnamefont
  {S.~J.}\ \bibnamefont {Dai}},\ and\ \bibinfo {author} {\bibfnamefont {F.~R.}\
  \bibnamefont {Xu}},\ }\href {https://doi.org/10.1016/j.physletb.2017.03.054}
  {\bibfield  {journal} {\bibinfo  {journal} {Phys. Lett. B}\ }\textbf
  {\bibinfo {volume} {769}},\ \bibinfo {pages} {227} (\bibinfo {year}
  {2017})}\BibitemShut {NoStop}%
\bibitem [{\citenamefont {Feshbach}(1958)}]{Feshbach1958}%
  \BibitemOpen
  \bibfield  {author} {\bibinfo {author} {\bibfnamefont {H.}~\bibnamefont
  {Feshbach}},\ }\href {https://doi.org/10.1146/annurev.ns.08.120158.000405}
  {\bibfield  {journal} {\bibinfo  {journal} {Annu. Rev. Nucl. Sci.}\ }\textbf
  {\bibinfo {volume} {8}},\ \bibinfo {pages} {49} (\bibinfo {year}
  {1958})}\BibitemShut {NoStop}%
\bibitem [{\citenamefont {Nagamine}\ \emph {et~al.}(1970)\citenamefont
  {Nagamine}, \citenamefont {Uchida},\ and\ \citenamefont
  {Kobayashi}}]{Nagamine1970}%
  \BibitemOpen
  \bibfield  {author} {\bibinfo {author} {\bibfnamefont {K.}~\bibnamefont
  {Nagamine}}, \bibinfo {author} {\bibfnamefont {A.}~\bibnamefont {Uchida}},\
  and\ \bibinfo {author} {\bibfnamefont {S.}~\bibnamefont {Kobayashi}},\ }\href
  {https://doi.org/10.1016/0375-9474(70)90315-5} {\bibfield  {journal}
  {\bibinfo  {journal} {Nucl. Phys. A}\ }\textbf {\bibinfo {volume} {145}},\
  \bibinfo {pages} {203} (\bibinfo {year} {1970})}\BibitemShut {NoStop}%
\bibitem [{\citenamefont {Batty}(1971)}]{Batty1971}%
  \BibitemOpen
  \bibfield  {author} {\bibinfo {author} {\bibfnamefont {C.~J.}\ \bibnamefont
  {Batty}},\ }\href {https://doi.org/10.1016/0375-9474(71)90181-3} {\bibfield
  {journal} {\bibinfo  {journal} {Nucl. Phys. A}\ }\textbf {\bibinfo {volume}
  {178}},\ \bibinfo {pages} {17} (\bibinfo {year} {1971})}\BibitemShut
  {NoStop}%
\bibitem [{\citenamefont {Fisher}\ \emph {et~al.}(1972)\citenamefont {Fisher},
  \citenamefont {Grench}, \citenamefont {Healey}, \citenamefont {McCarthy},
  \citenamefont {Parks},\ and\ \citenamefont {Whitney}}]{Fisher1972}%
  \BibitemOpen
  \bibfield  {author} {\bibinfo {author} {\bibfnamefont {T.~R.}\ \bibnamefont
  {Fisher}}, \bibinfo {author} {\bibfnamefont {H.~A.}\ \bibnamefont {Grench}},
  \bibinfo {author} {\bibfnamefont {D.~C.}\ \bibnamefont {Healey}}, \bibinfo
  {author} {\bibfnamefont {J.~S.}\ \bibnamefont {McCarthy}}, \bibinfo {author}
  {\bibfnamefont {D.}~\bibnamefont {Parks}},\ and\ \bibinfo {author}
  {\bibfnamefont {R.}~\bibnamefont {Whitney}},\ }\href
  {https://doi.org/10.1016/0375-9474(72)90367-3} {\bibfield  {journal}
  {\bibinfo  {journal} {Nucl. Phys. A}\ }\textbf {\bibinfo {volume} {179}},\
  \bibinfo {pages} {241} (\bibinfo {year} {1972})}\BibitemShut {NoStop}%
\bibitem [{\citenamefont {Birchall}\ \emph {et~al.}(1974)\citenamefont
  {Birchall}, \citenamefont {Conzett}, \citenamefont {Arvieux}, \citenamefont
  {Dahme},\ and\ \citenamefont {Larimer}}]{Birchall1974}%
  \BibitemOpen
  \bibfield  {author} {\bibinfo {author} {\bibfnamefont {J.}~\bibnamefont
  {Birchall}}, \bibinfo {author} {\bibfnamefont {H.~E.}\ \bibnamefont
  {Conzett}}, \bibinfo {author} {\bibfnamefont {J.}~\bibnamefont {Arvieux}},
  \bibinfo {author} {\bibfnamefont {W.}~\bibnamefont {Dahme}},\ and\ \bibinfo
  {author} {\bibfnamefont {R.~M.}\ \bibnamefont {Larimer}},\ }\href
  {https://doi.org/10.1016/0370-2693(74)90521-8} {\bibfield  {journal}
  {\bibinfo  {journal} {Phys. Lett. B}\ }\textbf {\bibinfo {volume} {53}},\
  \bibinfo {pages} {165} (\bibinfo {year} {1974})}\BibitemShut {NoStop}%
\bibitem [{\citenamefont {{von Przewoski}}\ \emph {et~al.}(1991)\citenamefont
  {{von Przewoski}}, \citenamefont {Eversheim}, \citenamefont {Hinterberger},
  \citenamefont {Lahr}, \citenamefont {Campbell}, \citenamefont {G\"otz},
  \citenamefont {Hammans}, \citenamefont {Henneck}, \citenamefont {Masson},
  \citenamefont {Sick},\ and\ \citenamefont {Bauhoff}}]{VonPrzewoski1991}%
  \BibitemOpen
  \bibfield  {author} {\bibinfo {author} {\bibfnamefont {B.}~\bibnamefont {{von
  Przewoski}}}, \bibinfo {author} {\bibfnamefont {P.~D.}\ \bibnamefont
  {Eversheim}}, \bibinfo {author} {\bibfnamefont {F.}~\bibnamefont
  {Hinterberger}}, \bibinfo {author} {\bibfnamefont {U.}~\bibnamefont {Lahr}},
  \bibinfo {author} {\bibfnamefont {J.}~\bibnamefont {Campbell}}, \bibinfo
  {author} {\bibfnamefont {J.}~\bibnamefont {G\"otz}}, \bibinfo {author}
  {\bibfnamefont {M.}~\bibnamefont {Hammans}}, \bibinfo {author} {\bibfnamefont
  {R.}~\bibnamefont {Henneck}}, \bibinfo {author} {\bibfnamefont
  {G.}~\bibnamefont {Masson}}, \bibinfo {author} {\bibfnamefont
  {I.}~\bibnamefont {Sick}},\ and\ \bibinfo {author} {\bibfnamefont
  {W.}~\bibnamefont {Bauhoff}},\ }\href
  {https://doi.org/10.1016/0375-9474(91)90422-3} {\bibfield  {journal}
  {\bibinfo  {journal} {Nucl. Phys. A}\ }\textbf {\bibinfo {volume} {528}},\
  \bibinfo {pages} {159} (\bibinfo {year} {1991})}\BibitemShut {NoStop}%
\bibitem [{\citenamefont {Henneck}\ \emph {et~al.}(1994)\citenamefont
  {Henneck}, \citenamefont {Masson}, \citenamefont {Eversheim}, \citenamefont
  {Gebel}, \citenamefont {Hinterberger}, \citenamefont {Lahr}, \citenamefont
  {Schmitt}, \citenamefont {Schleef},\ and\ \citenamefont {{von
  Przewoski}}}]{Henneck1994}%
  \BibitemOpen
  \bibfield  {author} {\bibinfo {author} {\bibfnamefont {R.}~\bibnamefont
  {Henneck}}, \bibinfo {author} {\bibfnamefont {G.}~\bibnamefont {Masson}},
  \bibinfo {author} {\bibfnamefont {P.~D.}\ \bibnamefont {Eversheim}}, \bibinfo
  {author} {\bibfnamefont {R.}~\bibnamefont {Gebel}}, \bibinfo {author}
  {\bibfnamefont {F.}~\bibnamefont {Hinterberger}}, \bibinfo {author}
  {\bibfnamefont {U.}~\bibnamefont {Lahr}}, \bibinfo {author} {\bibfnamefont
  {H.~W.}\ \bibnamefont {Schmitt}}, \bibinfo {author} {\bibfnamefont
  {J.}~\bibnamefont {Schleef}},\ and\ \bibinfo {author} {\bibfnamefont
  {B.}~\bibnamefont {{von Przewoski}}},\ }\href
  {https://doi.org/10.1016/0375-9474(94)90224-0} {\bibfield  {journal}
  {\bibinfo  {journal} {Nucl. Phys. A}\ }\textbf {\bibinfo {volume} {571}},\
  \bibinfo {pages} {541} (\bibinfo {year} {1994})}\BibitemShut {NoStop}%
\bibitem [{\citenamefont {Hannen}\ \emph {et~al.}(2003)\citenamefont {Hannen},
  \citenamefont {Amos}, \citenamefont {van~den Berg}, \citenamefont {Bieber},
  \citenamefont {Deb}, \citenamefont {Ellinghaus}, \citenamefont {Frekers},
  \citenamefont {Hagemann}, \citenamefont {Harakeh}, \citenamefont {Heyse},
  \citenamefont {de~Huu}, \citenamefont {Kr\"usemann}, \citenamefont {Rakers},
  \citenamefont {Schmidt}, \citenamefont {van~der Werf},\ and\ \citenamefont
  {W\"ortche}}]{Hannen2003}%
  \BibitemOpen
  \bibfield  {author} {\bibinfo {author} {\bibfnamefont {V.~M.}\ \bibnamefont
  {Hannen}}, \bibinfo {author} {\bibfnamefont {K.}~\bibnamefont {Amos}},
  \bibinfo {author} {\bibfnamefont {A.~M.}\ \bibnamefont {van~den Berg}},
  \bibinfo {author} {\bibfnamefont {R.~K.}\ \bibnamefont {Bieber}}, \bibinfo
  {author} {\bibfnamefont {P.~K.}\ \bibnamefont {Deb}}, \bibinfo {author}
  {\bibfnamefont {F.}~\bibnamefont {Ellinghaus}}, \bibinfo {author}
  {\bibfnamefont {D.}~\bibnamefont {Frekers}}, \bibinfo {author} {\bibfnamefont
  {M.}~\bibnamefont {Hagemann}}, \bibinfo {author} {\bibfnamefont {M.~N.}\
  \bibnamefont {Harakeh}}, \bibinfo {author} {\bibfnamefont {J.}~\bibnamefont
  {Heyse}}, \bibinfo {author} {\bibfnamefont {M.~A.}\ \bibnamefont {de~Huu}},
  \bibinfo {author} {\bibfnamefont {B.~A.~M.}\ \bibnamefont {Kr\"usemann}},
  \bibinfo {author} {\bibfnamefont {S.}~\bibnamefont {Rakers}}, \bibinfo
  {author} {\bibfnamefont {R.}~\bibnamefont {Schmidt}}, \bibinfo {author}
  {\bibfnamefont {S.~Y.}\ \bibnamefont {van~der Werf}},\ and\ \bibinfo {author}
  {\bibfnamefont {H.~J.}\ \bibnamefont {W\"ortche}},\ }\href
  {https://doi.org/10.1103/PhysRevC.67.054321} {\bibfield  {journal} {\bibinfo
  {journal} {Phys. Rev. C}\ }\textbf {\bibinfo {volume} {67}},\ \bibinfo
  {pages} {054321} (\bibinfo {year} {2003})}\BibitemShut {NoStop}%
\bibitem [{\citenamefont {Nagadi}\ \emph {et~al.}(2004)\citenamefont {Nagadi},
  \citenamefont {Weisel}, \citenamefont {Walter}, \citenamefont {Delaroche},\
  and\ \citenamefont {Romain}}]{Nagadi2004}%
  \BibitemOpen
  \bibfield  {author} {\bibinfo {author} {\bibfnamefont {M.~M.}\ \bibnamefont
  {Nagadi}}, \bibinfo {author} {\bibfnamefont {G.~J.}\ \bibnamefont {Weisel}},
  \bibinfo {author} {\bibfnamefont {R.~L.}\ \bibnamefont {Walter}}, \bibinfo
  {author} {\bibfnamefont {J.~P.}\ \bibnamefont {Delaroche}},\ and\ \bibinfo
  {author} {\bibfnamefont {P.}~\bibnamefont {Romain}},\ }\href
  {https://doi.org/10.1103/PhysRevC.70.064604} {\bibfield  {journal} {\bibinfo
  {journal} {Phys. Rev. C}\ }\textbf {\bibinfo {volume} {70}},\ \bibinfo
  {pages} {064604} (\bibinfo {year} {2004})}\BibitemShut {NoStop}%
\bibitem [{\citenamefont {Cunningham}\ \emph {et~al.}(2013)\citenamefont
  {Cunningham}, \citenamefont {Al-Khalili},\ and\ \citenamefont
  {Johnson}}]{Cunningham2013}%
  \BibitemOpen
  \bibfield  {author} {\bibinfo {author} {\bibfnamefont {E.~S.}\ \bibnamefont
  {Cunningham}}, \bibinfo {author} {\bibfnamefont {J.~S.}\ \bibnamefont
  {Al-Khalili}},\ and\ \bibinfo {author} {\bibfnamefont {R.~C.}\ \bibnamefont
  {Johnson}},\ }\href {https://doi.org/10.1103/PhysRevC.87.054601} {\bibfield
  {journal} {\bibinfo  {journal} {Phys. Rev. C}\ }\textbf {\bibinfo {volume}
  {87}},\ \bibinfo {pages} {054601} (\bibinfo {year} {2013})}\BibitemShut
  {NoStop}%
\bibitem [{\citenamefont {Khoa}\ \emph {et~al.}(2022)\citenamefont {Khoa},
  \citenamefont {Tan},\ and\ \citenamefont {Khoa}}]{Khoa2022}%
  \BibitemOpen
  \bibfield  {author} {\bibinfo {author} {\bibfnamefont {N.~H.~D.}\
  \bibnamefont {Khoa}}, \bibinfo {author} {\bibfnamefont {N.~H.}\ \bibnamefont
  {Tan}},\ and\ \bibinfo {author} {\bibfnamefont {D.~T.}\ \bibnamefont
  {Khoa}},\ }\href {https://doi.org/10.1103/PhysRevC.105.065802} {\bibfield
  {journal} {\bibinfo  {journal} {Phys. Rev. C}\ }\textbf {\bibinfo {volume}
  {105}},\ \bibinfo {pages} {065802} (\bibinfo {year} {2022})}\BibitemShut
  {NoStop}%
\bibitem [{\citenamefont {Vautherin}\ and\ \citenamefont
  {V\'en\'eroni}(1968)}]{vautherin68}%
  \BibitemOpen
  \bibfield  {author} {\bibinfo {author} {\bibfnamefont {D.}~\bibnamefont
  {Vautherin}}\ and\ \bibinfo {author} {\bibfnamefont {M.}~\bibnamefont
  {V\'en\'eroni}},\ }\href {https://doi.org/10.1016/0370-2693(68)90410-3}
  {\bibfield  {journal} {\bibinfo  {journal} {Phys. Lett. B}\ }\textbf
  {\bibinfo {volume} {26}},\ \bibinfo {pages} {552} (\bibinfo {year}
  {1968})}\BibitemShut {NoStop}%
\bibitem [{\citenamefont {Dover}\ and\ \citenamefont {{Van
  Giai}}(1971)}]{dover71}%
  \BibitemOpen
  \bibfield  {author} {\bibinfo {author} {\bibfnamefont {C.~B.}\ \bibnamefont
  {Dover}}\ and\ \bibinfo {author} {\bibfnamefont {N.}~\bibnamefont {{Van
  Giai}}},\ }\href {https://doi.org/10.1016/0375-9474(71)90308-3} {\bibfield
  {journal} {\bibinfo  {journal} {Nucl. Phys. A}\ }\textbf {\bibinfo {volume}
  {177}},\ \bibinfo {pages} {559} (\bibinfo {year} {1971})}\BibitemShut
  {NoStop}%
\bibitem [{\citenamefont {Dover}\ and\ \citenamefont {{Van
  Giai}}(1972)}]{dover72}%
  \BibitemOpen
  \bibfield  {author} {\bibinfo {author} {\bibfnamefont {C.~B.}\ \bibnamefont
  {Dover}}\ and\ \bibinfo {author} {\bibfnamefont {N.}~\bibnamefont {{Van
  Giai}}},\ }\href {https://doi.org/10.1016/0375-9474(72)90148-0} {\bibfield
  {journal} {\bibinfo  {journal} {Nucl. Phys. A}\ }\textbf {\bibinfo {volume}
  {190}},\ \bibinfo {pages} {373} (\bibinfo {year} {1972})}\BibitemShut
  {NoStop}%
\bibitem [{\citenamefont {{Le Anh}}\ and\ \citenamefont {{Minh
  Loc}}(2021)}]{anh2021PRC103}%
  \BibitemOpen
  \bibfield  {author} {\bibinfo {author} {\bibfnamefont {N.}~\bibnamefont {{Le
  Anh}}}\ and\ \bibinfo {author} {\bibfnamefont {B.}~\bibnamefont {{Minh
  Loc}}},\ }\href {https://doi.org/10.1103/PhysRevC.103.035812} {\bibfield
  {journal} {\bibinfo  {journal} {Phys. Rev. C}\ }\textbf {\bibinfo {volume}
  {103}},\ \bibinfo {pages} {035812} (\bibinfo {year} {2021})}\BibitemShut
  {NoStop}%
\bibitem [{\citenamefont {{Le Anh}}\ \emph {et~al.}(2021)\citenamefont {{Le
  Anh}}, \citenamefont {{Nhut Huan}},\ and\ \citenamefont {{Minh
  Loc}}}]{anh2021PRC104}%
  \BibitemOpen
  \bibfield  {author} {\bibinfo {author} {\bibfnamefont {N.}~\bibnamefont {{Le
  Anh}}}, \bibinfo {author} {\bibfnamefont {P.}~\bibnamefont {{Nhut Huan}}},\
  and\ \bibinfo {author} {\bibfnamefont {B.}~\bibnamefont {{Minh Loc}}},\
  }\href {https://doi.org/10.1103/PhysRevC.104.034622} {\bibfield  {journal}
  {\bibinfo  {journal} {Phys. Rev. C}\ }\textbf {\bibinfo {volume} {104}},\
  \bibinfo {pages} {034622} (\bibinfo {year} {2021})}\BibitemShut {NoStop}%
\bibitem [{\citenamefont {Le~Anh}\ and\ \citenamefont
  {Minh~Loc}(2022)}]{anh2022PRC}%
  \BibitemOpen
  \bibfield  {author} {\bibinfo {author} {\bibfnamefont {N.}~\bibnamefont
  {Le~Anh}}\ and\ \bibinfo {author} {\bibfnamefont {B.}~\bibnamefont
  {Minh~Loc}},\ }\href {https://doi.org/10.1103/PhysRevC.106.014605} {\bibfield
   {journal} {\bibinfo  {journal} {Phys. Rev. C}\ }\textbf {\bibinfo {volume}
  {106}},\ \bibinfo {pages} {014605} (\bibinfo {year} {2022})}\BibitemShut
  {NoStop}%
\bibitem [{\citenamefont {Ayyad}\ \emph {et~al.}(2022)\citenamefont {Ayyad},
  \citenamefont {Mittig}, \citenamefont {Tang}, \citenamefont {Olaizola},
  \citenamefont {Potel}, \citenamefont {Rijal}, \citenamefont {Watwood},
  \citenamefont {Alvarez-Pol}, \citenamefont {Bazin}, \citenamefont
  {Caama\~no}, \citenamefont {Chen}, \citenamefont {Cortesi}, \citenamefont
  {Fern\'andez-Dom\'{\i}nguez}, \citenamefont {Giraud}, \citenamefont {Gueye},
  \citenamefont {Heinitz}, \citenamefont {Jain}, \citenamefont {Kay},
  \citenamefont {Maugeri}, \citenamefont {Monteagudo}, \citenamefont
  {Ndayisabye}, \citenamefont {Paneru}, \citenamefont {Pereira}, \citenamefont
  {Rubino}, \citenamefont {Santamaria}, \citenamefont {Schumann}, \citenamefont
  {Surbrook}, \citenamefont {Wagner}, \citenamefont {Zamora},\ and\
  \citenamefont {Zelevinsky}}]{ayyad22}%
  \BibitemOpen
  \bibfield  {author} {\bibinfo {author} {\bibfnamefont {Y.}~\bibnamefont
  {Ayyad}}, \bibinfo {author} {\bibfnamefont {W.}~\bibnamefont {Mittig}},
  \bibinfo {author} {\bibfnamefont {T.}~\bibnamefont {Tang}}, \bibinfo {author}
  {\bibfnamefont {B.}~\bibnamefont {Olaizola}}, \bibinfo {author}
  {\bibfnamefont {G.}~\bibnamefont {Potel}}, \bibinfo {author} {\bibfnamefont
  {N.}~\bibnamefont {Rijal}}, \bibinfo {author} {\bibfnamefont
  {N.}~\bibnamefont {Watwood}}, \bibinfo {author} {\bibfnamefont
  {H.}~\bibnamefont {Alvarez-Pol}}, \bibinfo {author} {\bibfnamefont
  {D.}~\bibnamefont {Bazin}}, \bibinfo {author} {\bibfnamefont
  {M.}~\bibnamefont {Caama\~no}}, \bibinfo {author} {\bibfnamefont
  {J.}~\bibnamefont {Chen}}, \bibinfo {author} {\bibfnamefont {M.}~\bibnamefont
  {Cortesi}}, \bibinfo {author} {\bibfnamefont {B.}~\bibnamefont
  {Fern\'andez-Dom\'{\i}nguez}}, \bibinfo {author} {\bibfnamefont
  {S.}~\bibnamefont {Giraud}}, \bibinfo {author} {\bibfnamefont
  {P.}~\bibnamefont {Gueye}}, \bibinfo {author} {\bibfnamefont
  {S.}~\bibnamefont {Heinitz}}, \bibinfo {author} {\bibfnamefont
  {R.}~\bibnamefont {Jain}}, \bibinfo {author} {\bibfnamefont {B.~P.}\
  \bibnamefont {Kay}}, \bibinfo {author} {\bibfnamefont {E.~A.}\ \bibnamefont
  {Maugeri}}, \bibinfo {author} {\bibfnamefont {B.}~\bibnamefont {Monteagudo}},
  \bibinfo {author} {\bibfnamefont {F.}~\bibnamefont {Ndayisabye}}, \bibinfo
  {author} {\bibfnamefont {S.~N.}\ \bibnamefont {Paneru}}, \bibinfo {author}
  {\bibfnamefont {J.}~\bibnamefont {Pereira}}, \bibinfo {author} {\bibfnamefont
  {E.}~\bibnamefont {Rubino}}, \bibinfo {author} {\bibfnamefont
  {C.}~\bibnamefont {Santamaria}}, \bibinfo {author} {\bibfnamefont
  {D.}~\bibnamefont {Schumann}}, \bibinfo {author} {\bibfnamefont
  {J.}~\bibnamefont {Surbrook}}, \bibinfo {author} {\bibfnamefont
  {L.}~\bibnamefont {Wagner}}, \bibinfo {author} {\bibfnamefont {J.~C.}\
  \bibnamefont {Zamora}},\ and\ \bibinfo {author} {\bibfnamefont
  {V.}~\bibnamefont {Zelevinsky}},\ }\href
  {https://doi.org/10.1103/PhysRevLett.129.012501} {\bibfield  {journal}
  {\bibinfo  {journal} {Phys. Rev. Lett.}\ }\textbf {\bibinfo {volume} {129}},\
  \bibinfo {pages} {012501} (\bibinfo {year} {2022})}\BibitemShut {NoStop}%
\bibitem [{\citenamefont {Lopez-Saavedra}\ \emph {et~al.}(2022)\citenamefont
  {Lopez-Saavedra}, \citenamefont {Almaraz-Calderon}, \citenamefont {Asher},
  \citenamefont {Baby}, \citenamefont {Gerken}, \citenamefont {Hanselman},
  \citenamefont {Kemper}, \citenamefont {Kuchera}, \citenamefont {Morelock},
  \citenamefont {Perello}, \citenamefont {Temanson}, \citenamefont {Volya},\
  and\ \citenamefont {Wiedenh\"over}}]{lopez22}%
  \BibitemOpen
  \bibfield  {author} {\bibinfo {author} {\bibfnamefont {E.}~\bibnamefont
  {Lopez-Saavedra}}, \bibinfo {author} {\bibfnamefont {S.}~\bibnamefont
  {Almaraz-Calderon}}, \bibinfo {author} {\bibfnamefont {B.~W.}\ \bibnamefont
  {Asher}}, \bibinfo {author} {\bibfnamefont {L.~T.}\ \bibnamefont {Baby}},
  \bibinfo {author} {\bibfnamefont {N.}~\bibnamefont {Gerken}}, \bibinfo
  {author} {\bibfnamefont {K.}~\bibnamefont {Hanselman}}, \bibinfo {author}
  {\bibfnamefont {K.~W.}\ \bibnamefont {Kemper}}, \bibinfo {author}
  {\bibfnamefont {A.~N.}\ \bibnamefont {Kuchera}}, \bibinfo {author}
  {\bibfnamefont {A.~B.}\ \bibnamefont {Morelock}}, \bibinfo {author}
  {\bibfnamefont {J.~F.}\ \bibnamefont {Perello}}, \bibinfo {author}
  {\bibfnamefont {E.~S.}\ \bibnamefont {Temanson}}, \bibinfo {author}
  {\bibfnamefont {A.}~\bibnamefont {Volya}},\ and\ \bibinfo {author}
  {\bibfnamefont {I.}~\bibnamefont {Wiedenh\"over}},\ }\href
  {https://doi.org/10.1103/PhysRevLett.129.012502} {\bibfield  {journal}
  {\bibinfo  {journal} {Phys. Rev. Lett.}\ }\textbf {\bibinfo {volume} {129}},\
  \bibinfo {pages} {012502} (\bibinfo {year} {2022})}\BibitemShut {NoStop}%
\bibitem [{\citenamefont {Le~Anh}\ \emph {et~al.}(2022)\citenamefont {Le~Anh},
  \citenamefont {Minh~Loc}, \citenamefont {Auerbach},\ and\ \citenamefont
  {Zelevinsky}}]{anh2022Be10}%
  \BibitemOpen
  \bibfield  {author} {\bibinfo {author} {\bibfnamefont {N.}~\bibnamefont
  {Le~Anh}}, \bibinfo {author} {\bibfnamefont {B.}~\bibnamefont {Minh~Loc}},
  \bibinfo {author} {\bibfnamefont {N.}~\bibnamefont {Auerbach}},\ and\
  \bibinfo {author} {\bibfnamefont {V.}~\bibnamefont {Zelevinsky}},\ }\href
  {https://doi.org/10.1103/PhysRevC.106.L051302} {\bibfield  {journal}
  {\bibinfo  {journal} {Phys. Rev. C}\ }\textbf {\bibinfo {volume} {106}},\
  \bibinfo {pages} {L051302} (\bibinfo {year} {2022})}\BibitemShut {NoStop}%
\bibitem [{\citenamefont {Davies}\ and\ \citenamefont
  {Satchler}(1964)}]{davies64}%
  \BibitemOpen
  \bibfield  {author} {\bibinfo {author} {\bibfnamefont {K.~T.~R.}\
  \bibnamefont {Davies}}\ and\ \bibinfo {author} {\bibfnamefont {G.~R.}\
  \bibnamefont {Satchler}},\ }\href
  {https://doi.org/10.1016/0029-5582(64)90582-6} {\bibfield  {journal}
  {\bibinfo  {journal} {Nucl. Phys.}\ }\textbf {\bibinfo {volume} {53}},\
  \bibinfo {pages} {1} (\bibinfo {year} {1964})}\BibitemShut {NoStop}%
\bibitem [{\citenamefont {Amos}\ \emph {et~al.}(2003)\citenamefont {Amos},
  \citenamefont {Canton}, \citenamefont {Pisent}, \citenamefont {Svenne},\ and\
  \citenamefont {{van der Knijff}}}]{amos2003}%
  \BibitemOpen
  \bibfield  {author} {\bibinfo {author} {\bibfnamefont {K.}~\bibnamefont
  {Amos}}, \bibinfo {author} {\bibfnamefont {L.}~\bibnamefont {Canton}},
  \bibinfo {author} {\bibfnamefont {G.}~\bibnamefont {Pisent}}, \bibinfo
  {author} {\bibfnamefont {J.}~\bibnamefont {Svenne}},\ and\ \bibinfo {author}
  {\bibfnamefont {D.}~\bibnamefont {{van der Knijff}}},\ }\href
  {https://doi.org/10.1016/j.nuclphysa.2003.08.019} {\bibfield  {journal}
  {\bibinfo  {journal} {Nucl. Phys. A}\ }\textbf {\bibinfo {volume} {728}},\
  \bibinfo {pages} {65} (\bibinfo {year} {2003})}\BibitemShut {NoStop}%
\bibitem [{\citenamefont {Amos}\ \emph {et~al.}(2021)\citenamefont {Amos},
  \citenamefont {Fraser}, \citenamefont {Karataglidis},\ and\ \citenamefont
  {Canton}}]{amos21}%
  \BibitemOpen
  \bibfield  {author} {\bibinfo {author} {\bibfnamefont {K.}~\bibnamefont
  {Amos}}, \bibinfo {author} {\bibfnamefont {P.~R.}\ \bibnamefont {Fraser}},
  \bibinfo {author} {\bibfnamefont {S.}~\bibnamefont {Karataglidis}},\ and\
  \bibinfo {author} {\bibfnamefont {L.}~\bibnamefont {Canton}},\ }\href
  {https://doi.org/10.1140/epja/s10050-021-00479-8} {\bibfield  {journal}
  {\bibinfo  {journal} {Eur. Phys. J. A}\ }\textbf {\bibinfo {volume} {57}},\
  \bibinfo {pages} {165} (\bibinfo {year} {2021})}\BibitemShut {NoStop}%
\bibitem [{\citenamefont {Col{\`o}}\ \emph {et~al.}(2013)\citenamefont
  {Col{\`o}}, \citenamefont {Cao}, \citenamefont {{Van Giai}},\ and\
  \citenamefont {Capelli}}]{colo13}%
  \BibitemOpen
  \bibfield  {author} {\bibinfo {author} {\bibfnamefont {G.}~\bibnamefont
  {Col{\`o}}}, \bibinfo {author} {\bibfnamefont {L.}~\bibnamefont {Cao}},
  \bibinfo {author} {\bibfnamefont {N.}~\bibnamefont {{Van Giai}}},\ and\
  \bibinfo {author} {\bibfnamefont {L.}~\bibnamefont {Capelli}},\ }\href
  {https://doi.org/10.1016/j.cpc.2012.07.016} {\bibfield  {journal} {\bibinfo
  {journal} {Comput. Phys. Commun.}\ }\textbf {\bibinfo {volume} {184}},\
  \bibinfo {pages} {142} (\bibinfo {year} {2013})}\BibitemShut {NoStop}%
\bibitem [{\citenamefont {Raynal}(1971)}]{ECIS06}%
  \BibitemOpen
  \bibfield  {author} {\bibinfo {author} {\bibfnamefont {J.}~\bibnamefont
  {Raynal}},\ }\bibfield  {title} {\bibinfo {title} {{ECIS06 code is
  distributed by the NEA DATA Bank, Paris, France; ECIS72 described in
  “Optical model and coupled-channels calculations in nuclear physics,”}},\
  }in\ \href@noop {} {\emph {\bibinfo {booktitle} {Computing as a language of
  physics. ICTP International Seminar Course, Trieste, Italy}}}\ (\bibinfo
  {year} {1971})\ p.\ \bibinfo {pages} {281}\BibitemShut {NoStop}%
\bibitem [{\citenamefont {Ajzenberg-Selove}(1991)}]{selove91}%
  \BibitemOpen
  \bibfield  {author} {\bibinfo {author} {\bibfnamefont {F.}~\bibnamefont
  {Ajzenberg-Selove}},\ }\href {https://doi.org/10.1016/0375-9474(91)90446-D}
  {\bibfield  {journal} {\bibinfo  {journal} {Nucl. Phys. A}\ }\textbf
  {\bibinfo {volume} {523}},\ \bibinfo {pages} {1} (\bibinfo {year}
  {1991})}\BibitemShut {NoStop}%
\bibitem [{\citenamefont {Chabanat}\ \emph {et~al.}(1998)\citenamefont
  {Chabanat}, \citenamefont {Bonche}, \citenamefont {Haensel}, \citenamefont
  {Meyer},\ and\ \citenamefont {Schaeffer}}]{chabanat98}%
  \BibitemOpen
  \bibfield  {author} {\bibinfo {author} {\bibfnamefont {E.}~\bibnamefont
  {Chabanat}}, \bibinfo {author} {\bibfnamefont {P.}~\bibnamefont {Bonche}},
  \bibinfo {author} {\bibfnamefont {P.}~\bibnamefont {Haensel}}, \bibinfo
  {author} {\bibfnamefont {J.}~\bibnamefont {Meyer}},\ and\ \bibinfo {author}
  {\bibfnamefont {R.}~\bibnamefont {Schaeffer}},\ }\href
  {https://doi.org/10.1016/S0375-9474(98)00180-8} {\bibfield  {journal}
  {\bibinfo  {journal} {Nucl. Phys. A}\ }\textbf {\bibinfo {volume} {635}},\
  \bibinfo {pages} {231} (\bibinfo {year} {1998})}\BibitemShut {NoStop}%
\bibitem [{\citenamefont {Zhengmin}\ \emph {et~al.}(1993)\citenamefont
  {Zhengmin}, \citenamefont {Beijing}, \citenamefont {Zhenzhong},\ and\
  \citenamefont {Huimin}}]{liu93}%
  \BibitemOpen
  \bibfield  {author} {\bibinfo {author} {\bibfnamefont {L.}~\bibnamefont
  {Zhengmin}}, \bibinfo {author} {\bibfnamefont {L.}~\bibnamefont {Beijing}},
  \bibinfo {author} {\bibfnamefont {D.}~\bibnamefont {Zhenzhong}},\ and\
  \bibinfo {author} {\bibfnamefont {H.}~\bibnamefont {Huimin}},\ }\href
  {https://doi.org/10.1016/0168-583X(93)95977-D} {\bibfield  {journal}
  {\bibinfo  {journal} {Nucl. Instrum. Methods Phys. Res. B}\ }\textbf
  {\bibinfo {volume} {74}},\ \bibinfo {pages} {439} (\bibinfo {year}
  {1993})}\BibitemShut {NoStop}%
\bibitem [{\citenamefont {Meyer}(1976)}]{Meyer1976}%
  \BibitemOpen
  \bibfield  {author} {\bibinfo {author} {\bibfnamefont {H.~O.}\ \bibnamefont
  {Meyer}},\ }\href {https://doi.org/10.1007/BF01409090} {\bibfield  {journal}
  {\bibinfo  {journal} {Z. Physik A}\ }\textbf {\bibinfo {volume} {279}},\
  \bibinfo {pages} {41} (\bibinfo {year} {1976})}\BibitemShut {NoStop}%
\bibitem [{\citenamefont {Mazzoni}\ \emph {et~al.}(1998)\citenamefont
  {Mazzoni}, \citenamefont {Chiari}, \citenamefont {Giuntini}, \citenamefont
  {Mand\`o},\ and\ \citenamefont {Taccetti}}]{mazzoni98}%
  \BibitemOpen
  \bibfield  {author} {\bibinfo {author} {\bibfnamefont {S.}~\bibnamefont
  {Mazzoni}}, \bibinfo {author} {\bibfnamefont {M.}~\bibnamefont {Chiari}},
  \bibinfo {author} {\bibfnamefont {L.}~\bibnamefont {Giuntini}}, \bibinfo
  {author} {\bibfnamefont {P.}~\bibnamefont {Mand\`o}},\ and\ \bibinfo {author}
  {\bibfnamefont {N.}~\bibnamefont {Taccetti}},\ }\href
  {https://doi.org/10.1016/S0168-583X(97)00678-2} {\bibfield  {journal}
  {\bibinfo  {journal} {Nucl. Instrum. Methods Phys. Res. B}\ }\textbf
  {\bibinfo {volume} {136-138}},\ \bibinfo {pages} {86} (\bibinfo {year}
  {1998})}\BibitemShut {NoStop}%
\bibitem [{\citenamefont {Fortune}(1995)}]{fortune95}%
  \BibitemOpen
  \bibfield  {author} {\bibinfo {author} {\bibfnamefont {H.~T.}\ \bibnamefont
  {Fortune}},\ }\href {https://doi.org/10.1103/PhysRevC.52.2261} {\bibfield
  {journal} {\bibinfo  {journal} {Phys. Rev. C}\ }\textbf {\bibinfo {volume}
  {52}},\ \bibinfo {pages} {2261} (\bibinfo {year} {1995})}\BibitemShut
  {NoStop}%
\bibitem [{\citenamefont {Tamura}(1965)}]{Tamura1965}%
  \BibitemOpen
  \bibfield  {author} {\bibinfo {author} {\bibfnamefont {T.}~\bibnamefont
  {Tamura}},\ }\href {https://doi.org/10.1103/RevModPhys.37.679} {\bibfield
  {journal} {\bibinfo  {journal} {Rev. Mod. Phys.}\ }\textbf {\bibinfo {volume}
  {37}},\ \bibinfo {pages} {679} (\bibinfo {year} {1965})}\BibitemShut
  {NoStop}%
\bibitem [{\citenamefont {Stamp}(1967)}]{stamp67}%
  \BibitemOpen
  \bibfield  {author} {\bibinfo {author} {\bibfnamefont {A.~P.}\ \bibnamefont
  {Stamp}},\ }\href {https://doi.org/10.1103/PhysRev.153.1052} {\bibfield
  {journal} {\bibinfo  {journal} {Phys. Rev.}\ }\textbf {\bibinfo {volume}
  {153}},\ \bibinfo {pages} {1052} (\bibinfo {year} {1967})}\BibitemShut
  {NoStop}%
\bibitem [{\citenamefont {Hussein}\ and\ \citenamefont
  {Sherif}(1973)}]{Hussein1973}%
  \BibitemOpen
  \bibfield  {author} {\bibinfo {author} {\bibfnamefont {A.~H.}\ \bibnamefont
  {Hussein}}\ and\ \bibinfo {author} {\bibfnamefont {H.~S.}\ \bibnamefont
  {Sherif}},\ }\href {https://doi.org/10.1103/PhysRevC.8.518} {\bibfield
  {journal} {\bibinfo  {journal} {Phys. Rev. C}\ }\textbf {\bibinfo {volume}
  {8}},\ \bibinfo {pages} {518} (\bibinfo {year} {1973})}\BibitemShut {NoStop}%
\bibitem [{\citenamefont {McAbee}\ \emph {et~al.}(1990)\citenamefont {McAbee},
  \citenamefont {Thompson},\ and\ \citenamefont {Ohnishi}}]{MCABEE1990}%
  \BibitemOpen
  \bibfield  {author} {\bibinfo {author} {\bibfnamefont {T.~L.}\ \bibnamefont
  {McAbee}}, \bibinfo {author} {\bibfnamefont {W.~J.}\ \bibnamefont
  {Thompson}},\ and\ \bibinfo {author} {\bibfnamefont {H.}~\bibnamefont
  {Ohnishi}},\ }\href {https://doi.org/10.1016/0375-9474(90)90375-V} {\bibfield
   {journal} {\bibinfo  {journal} {Nucl. Phys. A}\ }\textbf {\bibinfo {volume}
  {509}},\ \bibinfo {pages} {39} (\bibinfo {year} {1990})}\BibitemShut
  {NoStop}%
\bibitem [{\citenamefont {Cunningham}\ \emph {et~al.}(2011)\citenamefont
  {Cunningham}, \citenamefont {Al-Khalili},\ and\ \citenamefont
  {Johnson}}]{Cunningham2011}%
  \BibitemOpen
  \bibfield  {author} {\bibinfo {author} {\bibfnamefont {E.~S.}\ \bibnamefont
  {Cunningham}}, \bibinfo {author} {\bibfnamefont {J.~S.}\ \bibnamefont
  {Al-Khalili}},\ and\ \bibinfo {author} {\bibfnamefont {R.~C.}\ \bibnamefont
  {Johnson}},\ }\href {https://doi.org/10.1103/PhysRevC.84.041601} {\bibfield
  {journal} {\bibinfo  {journal} {Phys. Rev. C}\ }\textbf {\bibinfo {volume}
  {84}},\ \bibinfo {pages} {041601} (\bibinfo {year} {2011})}\BibitemShut
  {NoStop}%
\bibitem [{\citenamefont {L\'epine-Szily}\ \emph {et~al.}(2003)\citenamefont
  {L\'epine-Szily}, \citenamefont {Oliveira}, \citenamefont {Galante},
  \citenamefont {Amadio}, \citenamefont {Vanin}, \citenamefont
  {Lichtenth\"aler}, \citenamefont {Guimar{\~a}es}, \citenamefont {Lima},
  \citenamefont {Bohlen}, \citenamefont {Ostrowski}, \citenamefont {{Di
  Pietro}}, \citenamefont {Laird}, \citenamefont {Maunoury}, \citenamefont {{de
  Oliveira Santos}}, \citenamefont {Roussel-Chomaz}, \citenamefont {Savajols},
  \citenamefont {Trinder}, \citenamefont {Villari},\ and\ \citenamefont {{de
  Vismes}}}]{lepine2003}%
  \BibitemOpen
  \bibfield  {author} {\bibinfo {author} {\bibfnamefont {A.}~\bibnamefont
  {L\'epine-Szily}}, \bibinfo {author} {\bibfnamefont {J.~M.}\ \bibnamefont
  {Oliveira}}, \bibinfo {author} {\bibfnamefont {D.}~\bibnamefont {Galante}},
  \bibinfo {author} {\bibfnamefont {G.}~\bibnamefont {Amadio}}, \bibinfo
  {author} {\bibfnamefont {V.}~\bibnamefont {Vanin}}, \bibinfo {author}
  {\bibfnamefont {R.}~\bibnamefont {Lichtenth\"aler}}, \bibinfo {author}
  {\bibfnamefont {V.}~\bibnamefont {Guimar{\~a}es}}, \bibinfo {author}
  {\bibfnamefont {G.~F.}\ \bibnamefont {Lima}}, \bibinfo {author}
  {\bibfnamefont {H.~G.}\ \bibnamefont {Bohlen}}, \bibinfo {author}
  {\bibfnamefont {A.~N.}\ \bibnamefont {Ostrowski}}, \bibinfo {author}
  {\bibfnamefont {A.}~\bibnamefont {{Di Pietro}}}, \bibinfo {author}
  {\bibfnamefont {A.~M.}\ \bibnamefont {Laird}}, \bibinfo {author}
  {\bibfnamefont {L.}~\bibnamefont {Maunoury}}, \bibinfo {author}
  {\bibfnamefont {F.}~\bibnamefont {{de Oliveira Santos}}}, \bibinfo {author}
  {\bibfnamefont {P.}~\bibnamefont {Roussel-Chomaz}}, \bibinfo {author}
  {\bibfnamefont {H.}~\bibnamefont {Savajols}}, \bibinfo {author}
  {\bibfnamefont {W.}~\bibnamefont {Trinder}}, \bibinfo {author} {\bibfnamefont
  {A.~C.~C.}\ \bibnamefont {Villari}},\ and\ \bibinfo {author} {\bibfnamefont
  {A.}~\bibnamefont {{de Vismes}}},\ }\href
  {https://doi.org/https://doi.org/10.1016/S0375-9474(03)01418-0} {\bibfield
  {journal} {\bibinfo  {journal} {Nucl. Phys. A}\ }\textbf {\bibinfo {volume}
  {722}},\ \bibinfo {pages} {C512} (\bibinfo {year} {2003})}\BibitemShut
  {NoStop}%
\end{thebibliography}%

\end{document}